\UseRawInputEncoding
\documentclass[preprint,aps,12pt,showpacs,nofootinbib,tightenlines]{revtex4}
\usepackage{mathrsfs}
\usepackage{amsmath}
\usepackage{amssymb}
\usepackage{epsfig}
\usepackage{epstopdf}
\usepackage{graphicx}
\usepackage{subfigure}
\usepackage{booktabs}
\usepackage{float}
\usepackage{amsmath}
\usepackage{color}
\usepackage{multirow}
\usepackage{bm}

\def\be{\begin{eqnarray}}
\def\en{\end{eqnarray}}
\def\non{\nonumber\\}

\begin{document}
\title{Semileptonic and nonleptonic weak decays of $\psi(1S,2S)$ and $\eta_{c}(1S,2S)$ to $D_{(s)}$ in the covariant light-Front approach}
\author{Zhi-Jie Sun, Zhi-Qing Zhang
\footnote{Electronic address: zhangzhiqing@haut.edu.cn}, You-Ya Yang and Hao Yang  } 
\affiliation{\it \small  Institute of Theoretical Physics, School of Sciences, Henan University of Technology,
 Zhengzhou, Henan 450052, China } 
\date{\today}
\begin{abstract}
In addition to the strong and electromagnetic decay modes,  the $\psi(1S,2S)$ and $\eta_{c}(1S,2S)$ can also decay via the weak interaction.
Such weak decays can be detected by the high-luminosity heavy-flavor experiments. At present, some of the semileptonic and nonleptonic $J/\Psi$ weak decays have been measured at BESIII. Researching for these charmonium weak decays to
$D_{(s)}$ meson can provide a platform to check of the standard model (SM) and probe new physics (NP). So we investigate the semileptonic and nonleptonic weak decays of $\psi(1S,2S)$ and $\eta_{c}(1S,2S)$ to $D_{(s)}$ within the covariant light-front quark model (CLFQM). With
form factors of the transitions $\psi(1S,2S)\to D_{(s)}$ and $\eta_{c}(1S,2S)\to D_{(s)}$ calculated under the CLFQM, we predict and discuss some physical observables, such as the branching ratios,  the longitudinal polarizations $f_{L}$ and the forward-backward asymmetries $A_{FB}$. One can find that the Cabibbo-favored semi-leptonic decay channels $\psi(1S,2S)\to D_{s}^{-}\ell^{+}\nu_{\ell}$ with $\ell=e,\mu$ and the nonleptonic decay modes $\psi(1S,2S)\to D_{s}^{-}\rho^{+}$ have relatively large branching ratios of the order $\mathcal{O}(10^{-9})$, which are most likely to be accessible at
the future high-luminosity experiments.
\end{abstract}

 \vspace{1cm}

\maketitle

\section{Introduction}\label{intro}
The $\psi(1S,2S)$ and $\eta_{c}(1S,2S)$ are S-wave charmonium states below the open-charm kinematic threshold. They predominantly decay through the strong and electromagnetic interactions. By contrast, their weak decays are rare processes due to the smallness of the weak interaction strength. While such decays have evoked a lot of
interest from  theoretical research \cite{k.R,Czarnecki,YMW,YMW2,Sun:2015bxp,Sun:2016ppe}, because they build a bridge between perturbative and nonperturbative physics and provide a valuable platform to comprehend the intricate behaviors and dynamics of strong interactions.
The hadronic decays of these charmonia via the annihilation of $c\bar c$ to gluons are of a high order in strong coupling $\alpha_s$ and are severely
suppressed by the phenomenological Okubo-Zweig-Iizuka (OZI) rule \cite{Okubo:1963fa,CERN,Iizuka:1966fk}. Numerically£¬ the total branching ratio of the charmoium weak decays was estimated to be at the order of $10^{-8}$ \cite{Sanchis-Lozano:1993vyw}. New physics may have a chance to show up in such rare decays. Furthermore, for the weak decays of charmonia $\psi(1S,2S)$, the polarization effect may play an important role to probe the underlying dynamics and hadron structures \cite{Sanchis-Lozano:1993vyw}.

The BESIII Collaboration has reported on the results of searches for the hadronic and semileptonic weak decays $J/\psi\to D^{-}_{s}\pi^{+}$, $J/\psi\to D^{-}\pi^{+}, J/\psi\to \bar D^0 \bar K^0$ \cite{BES:2007hqc}, $J/\psi\to D^{-}_{s}\rho^{+}$\cite{BESIII:2014xbo}, $J/\psi\to D_{s}^{(\ast)-}e^{+}\nu_{e}$ \cite{Wang:2022ghh}, $J/\psi\to D^{-}e^{+}\nu_{e}$ \cite{BESIII:2021mnd}, respectively. Very recently, the semileptonic weak decay  $J/\psi\rightarrow D^{-}\mu^{+}\nu_{\mu}$ was firstly researched at BESIII \cite{BESIII:2023fqz}. The branching ratios at $90\%$ confidence level were found to be $\mathcal{B}r(J/\psi\to D^{-}_{s}\pi^{+})<1.4\times10^{-4}$, $\mathcal{B}r(J/\psi\to D^{-}\pi^{+})<7.5\times10^{-5}, \mathcal{B}r(J/\psi\to \bar D^0 \bar K^0)<1.7\times10^{-4}$, $\mathcal{B}r(J/\psi\to D^{-}_{s}\rho^{+})<1.3\times10^{-5}$, $\mathcal{B}r(J/\psi\rightarrow D_{s}^{(\ast)-}e^{+}\nu_{e})<1.3\times10^{-6}$, $\mathcal{B}r(J/\psi\rightarrow D^{-}e^{+}\nu_{e})<7.1\times10^{-8}$ and $\mathcal{B}r(J/\psi\rightarrow D^{-}\mu^{+}\nu_{\mu})<5.6\times10^{-7}$. Certainly, these upper limits greatly exceed the predicted values within the Standard Model (SM), which are in the order of $10^{-9}\sim 10^{-12}$ \cite{Czarnecki,Sun:2015bxp,R.R,Shen:2008zzb,Y.M,k.R,Sun:2016ppe,Wang:2007ys,Sun:2015nra,Wang:2016dkd,Yang:2016gnh,Ivanov:2015woa}. Even so, with the significant annual accumulation of $10^{10} J/\psi$ events, BESIII will soon be capable of detecting some of these decays in the near future.

For the semileptonic decays, the hadronic transition matrix element between the initial and final mesons is most crucial for the theoretical calculations, which
can be characterized by several form factors. As to the form factors, they can be extracted from data or relied on
some non-perturbative methods. The covariant light-front quark model (CLFQM) as one of popular non-perturbative methods has been successfully used to calculate the form factors \cite{Y. Cheng,H.Y. Cheng,Hwangw,CDL,WYC,Zhang:2023ypl}.
 Compared with the semileptonic decays, the nonleptonic decays are more complex in dynamics due to both of the two final states being hadrons, where more long distance effects are involved. The factorization
 assumption based on the vacuum saturation approximation is often used to simplify the calculations. Specificly, the matrix elements are factorized into a
 product of two single matrix elements of currents, where one is parameterized by the decay constant of the emitted meson and the other is represented by
the transition form factor. In a word, the form factors are important to both semileptonic and nonleptonic decays.
 A variety of models have been applied to study the transition form factors, such as the Bauer-Stech-Wirbel (BSW) model \cite{R.R}, the QCD sum rules (QCDSR) \cite{YMW,YMW2,Y.M}, the Bethe-Salpeter (BS) method \cite{Wang:2016dkd}. Based on the form factors and helicity formalisms, we also calculate another two physical observables: the forward-backward asymmetry $A_{FB}$ and the longitudinal polarization fraction $f_{L}$, respectively.

This paper is organized as follows. The formalism of the CLFQM, the hadronic matrix elements and the helicity amplitudes combined via form factors are listed in Sec. \ref{form1}. In addition to the numerical results for the
$\psi(1S,2S)\to D_{(s)}$ and $\eta_{c}(1S,2S)\to D_{(s)}$ transition form factors, the branching ratios, the forward-backward asymmetries $A_{FB}$ and the longitudinal polarization fractions $f_{L}$ for the corresponding decays are presented in Sec. \ref{numer}. Detailed comparisons with other theoretical values and relevant discussions are also included. The summary is presented in Sec. \ref{sum}. Some specific rules when performing the $p^-$ integration and the expression for each form factor are collected in the Appendix A and B, respectively.
\section{Formalism}\label{form1}
\subsection{The form factors}
The Bauer-Stech-Wirble (BSW) form factors for the $\eta_{c} \rightarrow D_{(s)}$ and  $J/\psi \rightarrow D_{(s)}$ transitions are defined as follows \footnote{It is similar for the $\eta_{c}(2S) \rightarrow D_{(s)}$ and  $\psi(2S) \rightarrow D_{(s)}$ transitions},
\begin{footnotesize}
\be
\left\langle D_{(s)}\left(P^{\prime
\prime}\right)\left|V_{\mu}\right|
\eta_{c}\left(P^{\prime}\right)\right\rangle
&=&\left(P_{\mu}-\frac{m_{\eta_{c}}^{2}-m_{D_{(s)}}^{2}}{q^{2}}
q_{\mu}\right) F_{1}^{\eta_{c}
D_{(s)}}\left(q^{2}\right)+\frac{m_{\eta_{c}}^{2}-m_{D_{(s)}}^{2}}{q^{2}} q_{\mu}
F_{0}^{\eta_{c} D_{(s)}}\left(q^{2}\right),\non \\
\left\langle D_{(s)}\left(P^{\prime \prime}\right)\left|V_{\mu}-A_{\mu}\right| J / \psi \left(P^{\prime}, \epsilon\right)\right\rangle
&=&-\epsilon_{\mu \nu \alpha \beta} \epsilon_{J / \psi}^{\nu} q^{\alpha} P^{\beta} \frac{V\left(q^{2}\right)}{m_{J / \psi}+m_{D_{(s)}}}-i \frac{2 m_{J / \psi} \epsilon_{J / \psi} \cdot q}{q^{2}} q_{\mu} A_{0}\left(q^{2}\right) \non
&&-i \epsilon_{J / \psi, \mu}\left(m_{J / \psi}+m_{D_{(s)}}\right) A_{1}\left(q^{2}\right)-i \frac{\epsilon_{J / \psi} \cdot q}{m_{ J / \psi}+m_{D_{(s)}}}P_{\mu} A_{2}\left(q^{2}\right) \non
&&+i \frac{2 m_{J / \psi} \epsilon_{J / \psi} \cdot q}{q^{2}} q_{\mu} A_{3}\left(q^{2}\right),
\en
\end{footnotesize}
where $P=P'+P'', q=P'-P''$ and the convention $\epsilon_{0123}=1$ is adopted. In order to calculate the amplitudes of the transition form factors, we need the following Feynman rules for the meson-quark-antiquark vertex $i\Gamma^{'} _{M}$, where the subscript $M$ represents a pseudoscalar (P) and
vector (V) scalar meson
\be
i\Gamma^{'} _{P}&=&H^{'}_{P}\gamma_{5},\\
i\Gamma^{'} _{V}&=&i H_{V}^{\prime}\left[\gamma_{\mu}-\frac{1}{W_{V}^{\prime}}\left(p_{1}^{\prime}-p_{2}\right)_{\mu}\right].
\en

\begin{figure}[htbp]
\centering \subfigure{
\begin{minipage}{5cm}
\centering
\includegraphics[width=5cm]{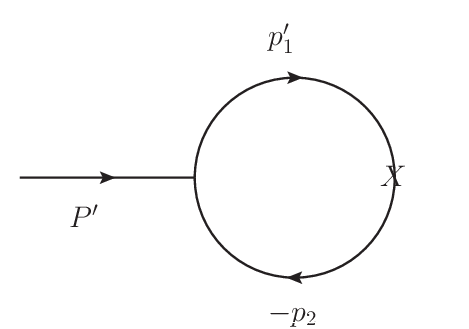}
\end{minipage}}
\subfigure{
\begin{minipage}{6cm}
\centering
\includegraphics[width=6cm]{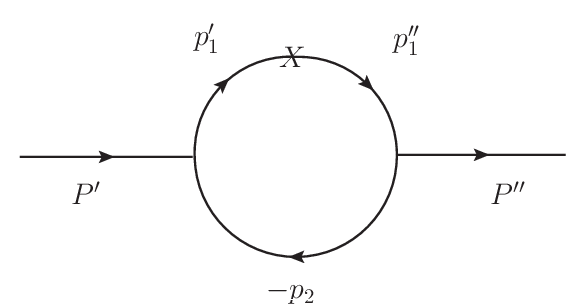}
\end{minipage}}
\caption{Feynman diagrams for charmonium decay (left) and transition
(right) amplitudes, where $P^{\prime(\prime\prime)}$ is the
incoming (outgoing) meson momentum, $p^{\prime(\prime\prime)}_1$
is the quark momentum, $p_2$ is the anti-quark momentum and X
denotes the vector or axial-vector transition vertex.}
\label{feyn}
\end{figure}

The results of the lowest order form factor could be obtained by calculating the Feynman diagram shown in Figure \ref{feyn}, where the Feynman diagram for the charmonium decay is also included. In the covariant quark model, the treatment of transition form factor is relatively covariant throughout the calculation process, where the light-front coordinates of a momentum $p$ are used $p=(p^-,p^+,p_\perp)$ with
$p^\pm=p^0\pm p_z, p^2=p^+p^--p^2_\perp$.
The incoming (outgoing) meson has the mass $M^\prime$ $(M^{\prime\prime})$
with the momentum $P^\prime=p_1^\prime+p_2$ $ (P^{\prime\prime}=p_1^{\prime\prime}+p_2)$, where $p_{1}^{\prime(\prime\prime)} $
and $p_{2}$ are the momenta of the quark and anti-quark
inside the incoming (outgoing) meson with the mass $m_{1}^{\prime(\prime\prime)}$and $m_{2}$, respectively. Here we use the same notations
as those in Refs. \cite{jaus,Y. Cheng} and $M^\prime$ refers to the charmonium mass.
These momenta can be expressed in terms of the internal variables $(x_{i},p{'}_{\perp})$ as
\be
p_{1,2}^{\prime+}=x_{1,2} P^{\prime+}, \quad p_{1,2 \perp}^{\prime}=x_{1,2} P_{\perp}^{\prime} \pm p_{\perp}^{\prime},
\en
with $x_{1}+x_{2}=1$  . Using these internal variables,
we can define some quantities for the incoming meson which will be used in the following calculations£»
\be
M_{0}^{\prime 2} &=&\left(e_{1}^{\prime}+e_{2}\right)^{2}=\frac{p_{\perp}^{\prime 2}+m_{1}^{\prime 2}}{x_{1}}
+\frac{p_{\perp}^{2}+m_{2}^{2}}{x_{2}}, \quad \widetilde{M}_{0}^{\prime}=\sqrt{M_{0}^{\prime 2}-\left(m_{1}^{\prime}-m_{2}\right)^{2}},\non
e_{i}^{(\prime)} &=&\sqrt{m_{i}^{(\prime) 2}+p_{\perp}^{\prime 2}+p_{z}^{\prime 2}}, \quad \quad p_{z}^{\prime}
=\frac{x_{2} M_{0}^{\prime}}{2}-\frac{m_{2}^{2}+p_{\perp}^{\prime 2}}{2 x_{2} M_{0}^{\prime}},
\en
where $M'_0$ is the kinetic invariant mass of the incoming meson and can be expressed as the energies of the quark and the anti-quark
$e^{(\prime)}_i$. It is similar to the case of the outgoing meson.
For the general $\eta_c(1S,2S)\to D_{(s)}$ transition, the decay amplitude for the lowest order is
\be
\mathcal{B}_{\mu}^{\eta_c D_{(s)}}=-i^{3} \frac{N_{c}}{(2 \pi)^{4}} \int d^{4} p_{1}^{\prime} \frac{H_{\eta_c}^{\prime}H_{D_{(s)}}^{\prime\prime}}
{N_{1}^{\prime} N_{1}^{\prime \prime} N_{2}} S_{\mu}^{\eta_c D_{(s)}}, \label{etacD}
\en
where $N_{1}^{\prime(\prime \prime)}=p_{1}^{\prime(\prime \prime) 2}-m_{1}^{\prime (\prime\prime) 2}$ and $N_{2}=p_{2}^{2}-m_{2}^{2} $ arise
from the quark propagators, and
the trace $ S_{\mu}^{\eta_c D_{(s)}}$ can be direct to obtain by using the Lorentz contraction,
\be
 S_{\mu}^{\eta_c D_{(s)}}&=&\operatorname{Tr}\left[\gamma_{5}\left(\not p_{1}^{\prime \prime}+m_{1}^{\prime \prime}\right) \gamma_{\mu}\left(\not p_{1}^{\prime}
+m_{1}^{\prime}\right) \gamma_{5}\left(-\not p_{2}+m_{2}\right)\right].
\label{ptop}
\en
It is similar for the $\psi(1S,2S)\to D_{(s)}$ transition amplitude,
\be
\mathcal{B}_{\mu}^{\psi D_{(s)}}=-i^{3} \frac{N_{c}}{(2 \pi)^{4}} \int d^{4} p_{1}^{\prime} \frac{H_{\psi}^{\prime}\left(i H_{D_{(s)}}^{\prime \prime}\right)}{N_{1}^{\prime} N_{1}^{\prime \prime} N_{2}}
 S_{\mu \nu}^{\psi D_{(s)}} \varepsilon^{*\nu},
\en
where
\be
S_{\mu \nu}^{\psi D_{(s)}}&=&\left(S_{V}^{\psi D_{(s)}}-S_{A}^{\psi D_{(s)}}\right)_{\mu \nu}\non
&=&\operatorname{Tr}\left[\left(\gamma_{\nu}-\frac{1}{W_{V}^{\prime \prime}}\left(p_{1}^{\prime \prime}-p_{2}\right)_{\nu}\right)\left(p_{1}^{\prime \prime}
+m_{1}^{\prime \prime}\right)\left(\gamma_{\mu}-\gamma_{\mu} \gamma_{5}\right)\left(\not p_{1}^{\prime}+m_{1}^{\prime}\right) \gamma_{5}\left(-\not p_{2}
+m_{2}\right)\right].\;\;\;\;\;\;
\label{sptov}
\en
The specific expressions for $S_{\mu}^{\eta_c D_{(s)}}$ and $S_{\mu \nu}^{\psi D_{(s)}}$ are listed in the Appendix B.
In practice, we use the light-front decomposition of the Feynman loop momentum and integrate out
the minus component through the contour method. If the covariant vertex functions are not singular when performing integration,
the transition amplitudes will
pick up the singularities in the anti-quark propagators. The integration then leads to
\be
N_{1}^{\prime(\prime \prime)} &\rightarrow& \hat{N}_{1}^{\prime(\prime \prime)}=x_{1}\left(M^{\prime(\prime \prime 2}-M_{0}^{\prime(\prime \prime) 2}\right),\non
H_{M}^{\prime(\prime\prime)} &\rightarrow& h_{M}^{\prime(\prime \prime)},\non
W_{M}^{\prime \prime} &\rightarrow& w_{M}^{\prime \prime}, \non
\int \frac{d^{4} p_{1}^{\prime}}{N_{1}^{\prime} N_{1}^{\prime \prime} N_{2}} H_{P}^{\prime} H_{M}^{\prime \prime} S^{PM} & \rightarrow&-i \pi \int \frac{d x_{2} d^{2}
p_{\perp}^{\prime}}{x_{2} \hat{N}_{1}^{\prime} \hat{N}_{1}^{\prime \prime}} h_{P}^{\prime} h_{M}^{\prime \prime} \hat{S}^{PM},
\en
where
\be
M_{0}^{\prime \prime 2}=\frac{p_{\perp}^{\prime \prime 2}+m_{1}^{\prime \prime 2}}{x_{1}}+\frac{p_{\perp}^{\prime \prime 2}+m_{2}^{2}}{x_{2}},
\label{vertex}
\en
with $p''_\perp=p'_\perp-x_2q_\perp$. The explicit forms of $h'_{M}$  and $w'_{M}$ are given by \cite{Y. Cheng}
\be
h_{P}^{\prime} &=&h_{V}^{\prime}=\left(M^{\prime 2}-M_{0}^{\prime 2}\right) \sqrt{\frac{x_{1} x_{2}}{N_{c}}} \frac{1}{\sqrt{2} \widetilde{M}_{0}^{\prime}} \varphi^{\prime},\\
w'_{V}&=&M^{'}_{0}+m^{'}_{1}+m_{2},\label{hp}
\en
with $\varphi^{\prime}$ being the light-front momentum distribution amplitude for the S-wave mesons,
\be
\varphi^{\prime} &=&\varphi^{\prime}\left(x_{2}, p_{\perp}^{\prime}\right)=4\left(\frac{\pi}{\beta^{\prime 2}}\right)^{\frac{3}{4}}
\sqrt{\frac{d p_{z}^{\prime}}{d x_{2}}} \exp \left(-\frac{p_{z}^{\prime 2}+p_{\perp}^{\prime 2}}{2 \beta^{\prime 2}}\right),
\en
where $\beta^\prime$ is a phenomenological parameter and can be fixed by fitting the corresponding decay constant. As to the radially excited charmonia $\psi(2S)$ and $\eta_c(2S)$, the distribution functions are given as
\be
\varphi^{\prime}(2S) &=&4\left(\frac{\pi}{\beta^{\prime 2}}\right)^{\frac{3}{4}}
\sqrt{\frac{d p_{z}^{\prime}}{d x_{2}}} \exp \left(-\frac{p_{z}^{\prime 2}+p_{\perp}^{\prime 2}}{2 \beta^{\prime 2}}\right)
\left(3-2\frac{p_{z}^{\prime 2}+p_{\perp}^{\prime 2}}{\beta^{\prime 2}}\right).
\en
Using Eqs. (\ref{etacD})-(\ref{vertex}) and taking the integration rules given in Refs \cite{jaus,Y. Cheng},
we can obtain the $\eta_c(1S,2S)\to D_{(s)}$ and $\psi(1S,2S)\to D_{(s)}$ transition form factors, which are shown in the Appendix B.
\subsection{Helicity amplitudes and Observables}\label{decaycons}
 Since the form factors involving the fitted parameters for the $\eta_{c}(1S,2S)\to D_{(s)}$ and $\psi(1S,2S)\to D_{(s)}$ transitions have been investigated  in above subsection, it is convenient to obtain the differential decay widths of these semileptontic $\eta_{c}(1S,2S)$ and $\psi(1S,2S)$ decays by the combination of the helicity amplitudes via form factors, which are listed as follows
 \begin{footnotesize}
\begin{eqnarray}
 \frac{d\Gamma(\eta_c\to D_{(s)}\ell\nu_\ell)}{dq^2} &=&(\frac{q^2-m_\ell^2}{q^2})^2\frac{ {\sqrt{\lambda(m_{\eta_c}^2,m_{D_{(s)}}^2,q^2)}} G_F^2 |V_{CKM}|^2} {384m_{\eta_c}^3\pi^3}
 \times \frac{1}{q^2} \nonumber\\
 &&\;\;\;\times \left\{ (m_\ell^2+2q^2) \lambda(m_{\eta_c}^2,m_{D_{(s)}}^2,q^2) F_1^2(q^2)  +3 m_\ell^2(m_{\eta_c}^2-m_{D_{(s)}}^2)^2F_0^2(q^2)
 \right\},\label{eq:pp}\\
 \frac{d\Gamma_L(\psi\to D_{(s)}\ell\nu_\ell)}{dq^2}&=&(\frac{q^2-m_\ell^2}{q^2})^2\frac{ {\sqrt{\lambda(m_{\psi}^2,m_{D_{(s)}}^2,q^2)}} G_F^2 |V_{CKM}|^2} {384m_{\psi}^3\pi^3}
 \times \frac{1}{q^2} \left\{ 3 m_\ell^2 \lambda(m_{\psi}^2,m_{D_{(s)}}^2,q^2) A_0^2(q^2)\right.\nonumber\\
 && +\frac{m_\ell^2+2q^2}{4m^2_{D_{(s)}}}\left.\left|
 (m_{\psi}^2-m_{D_{(s)}}^2-q^2)(m_{\psi}+m_{D_{(s)}})A_1(q^2)-\frac{\lambda(m_{\psi}^2,m_{D_{(s)}}^2,q^2)}{m_{\psi}+m_{D_{(s)}}}A_2(q^2)\right|^2
 \right\},\label{eq:decaywidthlon}\;\;\non\\
\frac{d\Gamma_\pm(\psi\to D_{(s)}\ell\nu_\ell)}{dq^2}&=&(\frac{q^2-m_\ell^2}{q^2})^2\frac{ {\sqrt{\lambda(m_{\psi}^2,m_{D_{(s)}}^2,q^2)}} G_F^2 |V_{CKM}|^2} {384m_{\psi}^3\pi^3}
  \nonumber\\
 &&\;\;\times \left\{ (m_\ell^2+2q^2) \lambda(m_{\psi}^2,m_{D_{(s)}}^2,q^2)\left|\frac{V(q^2)}{m_{\psi}+m_{D_{(s)}}}\mp
 \frac{(m_{\psi}+m_{D_{(s)}})A_1(q^2)}{\sqrt{\lambda(m_{\psi}^2,m_{D_{(s)}}^2,q^2)}}\right|^2
 \right\},\label{eq:widthlon2}
\end{eqnarray}
\end{footnotesize}
where $\lambda(q^2)=\lambda(m^{2}_{\eta_c(\psi)},m^{2}_{D_{(s)}},q^{2})=(m^{2}_{\eta_c(\psi)}+m^{2}_{D_{(s)}}-q^{2})^{2}-4m^{2}_{\eta_c(\psi)}m^{2}_{D_{(s)}}$ and $m_{\ell}$ is the mass of the lepton $\ell$ with $\ell=e,\mu$ \footnote{For now on, we use $\ell$ to represent $e,\mu$ for simplicity. }. It is noted that although the electron and nuon are very light compared with the charm quark, we do not ignore their masses in our calculations in order to check the mass effects.
The combined transverse and total differential decay widths are defined as
\be
\frac{d \Gamma_{T}}{d q^{2}}=\frac{d \Gamma_{+}}{d q^{2}}+\frac{d \Gamma_{-}}{d q^{2}}, \quad \frac{d \Gamma}{d q^{2}}=\frac{d \Gamma_{L}}{d q^{2}}+\frac{d \Gamma_{T}}{d q^{2}}.
\en

For the $\psi(1S,2S)$ decays, it is meaningful to define the polarization fraction due to the existence of different polarizations
\be
f_{L}=\frac{\Gamma_{L}}{\Gamma_{L}+\Gamma_{+}+\Gamma_{-}}. \label{eq:fl}
\en
As to the forward-backward asymmetry, the analytical expression is defined as \cite{Sakaki:2013bfa}
\be
A_{FB} = \frac{\int^1_0 {d\Gamma \over dcos\theta} dcos\theta - \int^0_{-1} {d\Gamma \over dcos\theta} dcos\theta}
{\int^1_{-1} {d\Gamma \over dcos\theta} dcos\theta} = \frac{\int b_\theta(q^2) dq^2}{\Gamma_{\eta_{c}(\psi)}},\label{eq:AFB}
\en
where $\theta$ is the angle between the 3-momenta of the lepton $\ell$ and the initial meson in the $\ell\nu$ rest frame. The function $b_{\theta}(q^2)$ represents the angular coefficient, which can be written as \cite{Sakaki:2013bfa}
\be
b_\theta(q^2) &=& {G_F^2 |V_{CKM}|^2 \over 128\pi^3 m_{\eta_{c}}^3} q^2 \sqrt{\lambda(q^2)}
\left( 1 - {m_\ell^2 \over q^2} \right)^2 {m_\ell^2 \over q^2} ( H^s_{V,0}H^s_{V,t} ) , \label{eq:btheta1}\\
b_\theta(q^2) &=& {G_F^2 |V_{CKM}|^2 \over 128\pi^3 m_{\psi}^3} q^2 \sqrt{\lambda(q^2)}
\left( 1 - {m_\ell^2 \over q^2} \right)^2 \left[ {1 \over 2}(H_{V,+}^2-H_{V,-}^2)+ {m_\ell^2 \over q^2} ( H_{V,0}H_{V,t} ) \right],
\label{eq:btheta2}
\en
where the helicity amplitudes
\be
H^s_{V,0}\left(q^{2}\right)  =\sqrt{\frac{\lambda\left(q^{2}\right)}{q^{2}}} F_{1}\left(q^{2}\right),
H^s_{V,t}\left(q^{2}\right)  =\frac{m_{\eta_{c}}^{2}-m_{{D_{(S)}}}^{2}}{\sqrt{q^{2}}} F_{0}\left(q^{2}\right),
\en
for the $\eta_c(1S,2S)\to D_{(s)}$ transitions, and
the helicity amplitudes
\be
H_{V,\pm}\left(q^{2}\right)&=&\left(m_{\psi}+{m_{D_{(s)}}}\right) A_{1}\left(q^{2}\right) \mp \frac{\sqrt{\lambda\left(q^{2}\right)}}{m_{\psi}+m_{D_{(s)}}} V\left(q^{2}\right), \non
H_{V,0}\left(q^{2}\right)&=&\frac{m_{\psi}+m_{D_{(s)}}}{2 m_{\psi} \sqrt{q^{2}}}\left[-\left(m_{\psi}^{2}-m_{D_{(s)}}^{2}-q^{2}\right) A_{1}\left(q^{2}\right)+\frac{\lambda\left(q^{2}\right) A_{2}\left(q^{2}\right)}{\left(m_{\psi}+m_{D_{(s)}}\right)^{2}}\right],\non
H_{V,t}\left(q^{2}\right)&=&-\sqrt{\frac{\lambda\left(q^{2}\right)}{q^{2}}} A_{0}\left(q^{2}\right),
\en
for the $\psi(1S,2S)\to D_{(s)}$ transitions. Here the subscript $V$ in each helicity amplitude refers to the $\gamma_\mu(1-\gamma_5)$ current.
\subsection{Hadronic matrix elements}
In phenomenology, the effective Hamiltonian of charmomium weak decays $\psi(1S,2S)\to D_{(s)}M$ and $ \eta_c(1S,2S)\to D_{(s)}M$ with $M=\pi, K, \rho, K^*$ can be written as \cite{Buchalla}
\be
\mathcal{H}_{\mathrm{eff}}=\frac{G_{F}}{\sqrt{2}} \sum_{q_{1}, q_{2}} V_{c q_{1}}^{*} V_{u q_{2}}\left\{C_{1}(\mu) Q_{1}(\mu)+C_{2}(\mu) Q_{2}(\mu)\right\}+\text { H.c. }
\en
where $G_{F}$ is the Fermi coupling constant, $V^{\ast}_{cq_1}V_{uq_2}$ is the product of the CKM matrix elements with $q_{1(2)}=s, d$, and $C_{1,2}(\mu)$ are the Wilson coefficients.
The local tree four-quark operators $Q_{1,2}$ are defined by
\be
Q_{1}=\left[\bar{q}_{1, \alpha} \gamma_{\mu}\left(1-\gamma_{5}\right) c_{\alpha}\right]\left[\bar{u}_{\beta} \gamma^{\mu}\left(1-\gamma_{5}\right) q_{2, \beta}\right], \\
Q_{2}=\left[\bar{q}_{1, \alpha} \gamma_{\mu}\left(1-\gamma_{5}\right) c_{\beta}\right]\left[\bar{u}_{\beta} \gamma^{\mu}\left(1-\gamma_{5}\right) q_{2, \alpha}\right],
\en
where $\alpha$ and $\beta$ are color indices. Based on the effective Hamiltonian, the matrix elements for the decays $\eta_{c}(1S, 2S) \rightarrow D_{(s)}M$  can be expressed as
\be
\mathcal{A}\left(\eta_{c} \rightarrow D_{(s)} M\right)=\left\langle D_{(s)} M\left|\mathcal{H}_{\mathrm{eff}}\right| \eta_{c}\right\rangle=\frac{G_{F}}{\sqrt{2}} V_{c q_{1}}^{*} V_{u q_{2}} a_{1}\left\langle M\left|J^{\mu}\right| 0\right\rangle\left\langle D_{(s)}\left|J_{\mu}\right| \eta_{c}\right\rangle,
\en
where the combination of the Wilson coefficients $a_1=C_2+C_1/3$ and $\left\langle M\left|J^{\mu}\right| 0\right\rangle$ is defined as $\left\langle P(q)\left|A^{\mu}\right| 0\right\rangle=-if_Pq_\mu$ for pseudoscalar (P) mesons and $\left\langle V(q,\epsilon)\left|V^{\mu}\right| 0\right\rangle=f_Vm_V\epsilon^*_\mu$ for vector (V) mesons. Specifically, the total amplitude for each decay
channel can be further written as follows
\be
\mathcal{A}\left(\eta_{c} \rightarrow D_{s}^{-} \pi^{+}\right) & =&   i \frac{G_{F}}{\sqrt{2}}V_{u d} V_{c s}^{*} a_{1}\left(m_{\eta_{c}}^{2}-m_{D_{s}}^{2}\right) f_{\pi} F_{0}^{\eta_{c}D_{s}}\left(m_{\pi}^{2}\right) , \\
\mathcal{A}\left(\eta_{c} \rightarrow D_{s}^{-} K^{+}\right)  &=&  i \frac{G_{F}}{\sqrt{2}}V_{u s} V_{c s}^{*} a_{1}\left(m_{\eta_{c}}^{2}-m_{D_{s}}^{2}\right) f_{K} F_{0}^{\eta_{c}D_{s}}\left(m_{K}^{2}\right) , \\
\mathcal{A}\left(\eta_{c} \rightarrow D^{-} \pi^{+}\right) &=&  i \frac{G_{F}}{\sqrt{2}}V_{u d} V_{c d}^{*} a_{1}\left(m_{\eta_{c}}^{2}-m_{D}^{2}\right) f_{\pi} F_{0}^{\eta_{c}D}\left(m_{\pi}^{2}\right) , \\
\mathcal{A}\left(\eta_{c} \rightarrow D^{-} K^{+}\right)  &=&  i \frac{G_{F}}{\sqrt{2}}V_{u s} V_{c d}^{*} a_{1}\left(m_{\eta_{c}}^{2}-m_{D}^{2}\right) f_{K} F_{0}^{\eta_{c} D}\left(m_{K}^{2}\right) ,\\
\mathcal{A}\left(\eta_{c} \rightarrow D_{s}^{-} \rho^{+}\right)  &=&  \sqrt{2} G_{F}V_{u d} V_{c s}^{*} a_{1} m_{\rho}\left(\epsilon_{\rho}^{*} \cdot p_{\eta_{c}}\right) f_{\rho} F_{1}^{\eta_{c} D_{s}}\left(m_{\rho}^{2}\right) , \\
\mathcal{A}\left(\eta_{c} \rightarrow D_{s}^{-} K^{*+}\right) & =&  \sqrt{2} G_{F}V_{u s} V_{c s}^{*} a_{1} m_{K^{*}}\left(\epsilon_{K^{*}}^{*} \cdot p_{\eta_{c}}\right) f_{K^{*}} F_{1}^{\eta_{c}D_{s}}\left(m_{K^*}^{2}\right) , \\
\mathcal{A}\left(\eta_{c} \rightarrow D^{-} \rho^{+}\right) & =&  \sqrt{2} G_{F}V_{u d} V_{c d}^{*} a_{1} m_{\rho}\left(\epsilon_{\rho}^{*} \cdot p_{\eta_{c}}\right) f_{\rho} F_{1}^{\eta_{c}D}\left(m_{\rho}^{2}\right) , \\
\mathcal{A}\left(\eta_{c} \rightarrow D^{-} K^{*+}\right)  &=&  \sqrt{2} G_{F}V_{u s} V_{c d}^{*} a_{1} m_{K^{*}}\left(\epsilon_{K^{*}}^{*} \cdot p_{\eta_{c}}\right) f_{K^{*}} F_{1}^{\eta_{c} D}\left(m_{K^*}^{2}\right).
\en
In addition, the amplitudes for the decays $\psi(1S, 2S) \rightarrow D_{(s)}P$ with $P= \pi, K$ can be expressed as
\be
A( \psi \rightarrow D_{(s)} P)=\left\langle D_{(s)} P\left|\mathcal{H}_{e f f}\right| \psi\right\rangle=\frac{G_{F}}{\sqrt{2}} V_{c q_{1}}^{*} V_{u q_{2}} a_{1} 2 m_{\psi}\left(\epsilon_{\psi} \cdot p_P\right) f_{P} A^{\psi D_{(s)}}_{0}\left(m_{P}^{2}\right).
\en
As to the specific decay channels, the amplitudes are given as
\be
A\left( \psi \rightarrow D_{s}^{-} \pi^{+}\right) &=& \sqrt{2}G_{F} V_{u d} V_{c s}^{*} a_{1} m_{ \psi}\left(\epsilon_{\psi} \cdot p_\pi\right) f_{\pi} A_{0}^{ \psi D_{s}}\left(m_{\pi}^{2}\right),\\
A\left( \psi \rightarrow D_{s}^{-} K^{+}\right) &=&  \sqrt{2}G_{F} V_{u s} V_{c s}^{*} a_{1}m_{ \psi}\left(\epsilon_{\psi} \cdot p_K\right) f_{K} A_{0}^{\psi D_{s}}\left(m_{K}^{2}\right),\\
A\left( \psi \rightarrow D^{-} \pi^{+}\right) &=&  \sqrt{2}G_{F} V_{u d} V_{c d}^{*} a_{1}m_{ \psi} \left(\epsilon_{\psi} \cdot p_\pi\right)f_{\pi} A_{0}^{\psi D}\left(m_{\pi}^{2}\right),\\
A\left(\psi \rightarrow D^{-} K^{+}\right) &=&  \sqrt{2}G_{F} V_{u s} V_{c d}^{*} a_{1}m_{ \psi} \left(\epsilon_{\psi} \cdot p_K\right)f_{K} A_{0}^{\psi D}\left(m_{K}^{2}\right).
\en
For the decays $\psi(1S, 2S)\rightarrow D_{(s)} V$, the hadronic matrix elements can be expressed as
\be
\mathcal{A}\left( \psi \rightarrow D_{(s)} V \right)=\left\langle D_{(s)} V\left|\mathcal{H}_{\mathrm{eff}}\right|  \psi \right\rangle=\frac{G_{F}}{\sqrt{2}} V_{c q_{1}}^{*} V_{u q_{2}} a_{1} H_{\lambda},
\en
where $\lambda$ denotes the helicity of vector meson, and $\mathcal{H}_{\lambda}=\left\langle V\left|J^{\mu}\right| 0\right\rangle\left\langle P \left|J_{\mu}\right|  \psi\right\rangle$ is given as follows
\be
H_{0} &\equiv&  \left\langle V\left(\varepsilon_{0}^{\prime}, p_{V}\right)\left|\bar{q} \gamma^{\mu} q\right| 0\right\rangle\left\langle D_{(s)}\left(p_{D_{(s)}}\right)\left|\bar{c} \gamma_{\mu}\left(1-\gamma_{5}\right) b\right|  \psi\left(\varepsilon_{0}, p_{ \psi}\right)\right\rangle \non
&=&\frac{i f_{V}}{2 m_{ \psi}}\left[\left(m_{ \psi}^{2}-m_{D_{(s)}}^{2}+m_{V}^{2}\right)\left(m_{ \psi}+m_{D_{(s)}}\right) A_{1}^{ \psi D_{(s)}}\left(m_{V}^{2}\right)\right. \non&&
 \left.+\frac{4 m_{ \psi}^{2} p_{c}^{2}}{m_{ \psi}+m_{D_{(s)}}} A_{2}^{ \psi D_{(s)}}\left(m_{V}^{2}\right)\right], \\
H_{\pm} &\equiv& \left\langle V\left(\varepsilon_{\pm}^{\prime}, p_{V}\right)\left|\bar{q} \gamma^{\mu} q\right| 0\right\rangle\left\langle D_{(s)}\left(p_{D_{(s)}}\right)\left|\bar{c} \gamma_{\mu}\left(1-\gamma_{5}\right) b\right|  \psi\left(\varepsilon_{\pm}, p_{ \psi}\right)\right\rangle \non
&=&i f_{V} m_{V}\left[-\left(m_{\psi}+m_{D_{(s)}}\right) A_{1}^{\psi  D_{(s)}}\left(m_{V}^{2}\right) \pm \frac{2 m_{\psi} p_{c}}{m_{\psi}+m_{D}} V^{ \psi D_{(s)}}\left(m_{V}^{2}\right)\right].
\en
\section{Numerical results and discussions} \label{numer}
\subsection{Transition Form Factors}
\begin{table}[H]
\caption{The values of the input parameters \cite{Zhang:2023ypl,Workman,damir,chiu,Wingate}. }
\label{tab:constant}
\begin{tabular*}{16.5cm}{@{\extracolsep{\fill}}l|cccccc}
  \hline\hline
\textbf{Mass(\text{GeV})} &$m_{b}=4.8$
&$m_{c}=1.4$&$m_{s}=0.37$&$m_{u,d}=0.25$&$m_{e}=0.000511$   \\[1ex]
&$m_{\pi}=0.140$&$m_{K}=0.494$&$m_{\rho}=0.775$&$m_{K^{*}}=0.892$& $m_{\mu}=0.106$\\[1ex]
& $m_{\eta_c}=2.9839$& $ m_{J/\psi}=3.0969$  & $m_{\eta_c(2S)}=3.6377 $& $m_{\psi(2S)}=3.68610 $& $m_{D}=1.86966 $ \\[1ex]
& $m_{D_{s}}=1.96835 $ \\[1ex]
\hline
\end{tabular*}
\begin{tabular*}{16.5cm}{@{\extracolsep{\fill}}l|ccccc}
  \hline
\multirow{2}{*}{{\textbf{CKM}}}&$V_{cd}=0.221\pm0.004$&$V_{us}=0.2243\pm0.0008$\\[1ex]
& $V_{ud}=0.97373\pm0.00031$&$V_{cs}=0.975\pm0.006$ \\[1ex]
\hline
\end{tabular*}
\begin{tabular*}{16.5cm}{@{\extracolsep{\fill}}l|ccccc}
\hline
\textbf{ decay constants(\text{GeV})} & $f_{\pi}=0.132$ & $f_{K}=0.16$
& $f_{\rho}=0.209$  &$f_{K^{*}}=0.217$\\[1ex]
& $f_{J/\Psi}=0.431$ & $f_{\eta_c}=0.387$
& $f_{D}=0.235$  &$f_{D_s}=0.290$\\[1ex]
\hline\hline
\end{tabular*}
\begin{tabular*}{16.5cm}{@{\extracolsep{\fill}}l|ccccc}
\textbf{shape parameters(\text{GeV})}&$\beta^{'}_{\eta_{c}}=0.754^{+0.014}_{-0.014}$&$\beta^{'}_{\eta_{c}(2S)}=0.388^{+0.092}_{-0.096}$&$\beta^{'}_{D}=0.541^{+0.043}_{-0.042}$\\[1ex]
& $\beta^{'}_{J/\psi}=0.646^{+0.041}_{-0.041}$&$\beta^{'}_{\psi(2S)}=0.385^{+0.049}_{-0.068}$&$\beta^{'}_{D_{s}}=0.645^{+0.136}_{-0.117}$\\[1ex]
\hline\hline
\end{tabular*}
\begin{tabular*}{16.5cm}{@{\extracolsep{\fill}}l|ccc}
\textbf{Full width}&$\Gamma_{\eta_c}=(32.0\pm0.7)\text{MeV}$&$\Gamma_{J/\psi}=(92.6\pm1.7)\text{keV}$&$$ \\[1ex]
$$&$\Gamma_{\eta_c}(2S)=(11.3^{+3.2}_{-2.9})\text{MeV}$&$\Gamma_{\psi(2S)}=(294\pm8)\text{keV}$\\[1ex]
\hline\hline
\end{tabular*}
\end{table}
The input parameters, including the masses of the initial and the final mesons, the CKM matrix elements, the shape parameters fitted by the decay constants, the full widths of the initial mesons, and so on are listed in Table \ref{tab:constant}. It is noted that the decay constant of charmonium $\eta_c(2S)$ is calculated as following formula
\be
f_{\eta_{c}(2S)}=\sqrt{\frac{81 m_{\eta_{c}(2 S)} \Gamma_{\eta_{c}(2 S) \rightarrow \gamma \gamma}}{64 \pi \alpha_{e m}^{2}}},
\en
where $\Gamma_{\eta_{c}(2S)} \rightarrow \gamma \gamma=(1.3\pm0.6)$ keV is taken from the CLEO measurement \cite{asner}. Then one can obtain
$f_{\eta_c(2S)}=(189^{+40}_{-50})$ MeV with smaller uncertainty compared with $f_{\eta_c(2S)}=(243^{+79}_{-111})$ MeV \cite{Zhang:2023ypl}. As to the decay constant of $\psi(2S)$, it is estimated from the relation $\frac{f_{\psi(2S)}}{f_{J/\Psi}}=\frac{f_{\eta_c(2S)}}{f_{\eta_c}}$ \cite{zhongzhi} and given as $f_{\psi(2S)}=(210^{+43}_{-52})$ MeV.
Based on the input parameters from Table \ref{tab:constant}, one can obtain the numerical results of the transition form factors at $q^2=0$  shown in Table \ref{table2}.

All the computations are carried out within the $q^+=0$ reference frame, where the form factors can only be obtained at spacelike momentum transfers $q^2=-q^2_{\bot}\leq0$. It is need to know the form factors in the timelike region for the physical decay processes. Here we use the
following double-pole approximation to parametrize the form factors in the spacelike region and then extend to the timelike region,
\be
F\left(q^{2}\right)=\frac{F(0)}{1-a q^{2} / m^{2}+b q^{4} / m^{4}},
\en
where $m$ represents the initial meson mass and $F(q^{2})$ denotes the different form factors $F_{1},F_{0},V,A_{0},A_{1}$ and $A_{2}$.
The values of $a$ and $b$ can be obtained by performing a 3-parameter fit to the form factors in the range $-10 \text{GeV}^2\leq q^2\leq0$, which are collected in Table \ref{table2}. The uncertainties arise from the decay constants of the initial charmonia ($\eta_{c}(1S,2S),\psi(1S,2S)$) and the final charmed mesons ($D, D_{s}$).

\begin{table}[H]
\caption{Form factors of the transitions $\eta_{c}(1S,2S)\rightarrow D_{(s)}, \psi(1S,2S)\rightarrow D_{(s)}$ in the CLFQM.
The uncertainties are from the decay constants of initial and final state mesons.}
\begin{center}
\scalebox{0.9}{
\begin{tabular}{||ccc||cccc||}
\hline\hline
$$&$\eta_{c}\rightarrow D$&$$& $$&$J/\psi\rightarrow D $&$$&$$\\
\hline\hline
$$&$F_{1}$& $F_{0}$&$V$&$A_{0}$&$A_{1}$&$A_{2}$\\
\hline
$F(0)$&$0.73^{+0.00+0.03}_{-0.00-0.04}$&$0.73^{+0.00+0.03}_{-0.00-0.04}$&$1.73^{+0.01+0.03}_{-0.02-0.05}$&$0.45^{+0.02+0.02}_{-0.02-0.01}$&$0.53^{+0.01+0.00}_{-0.01-0.00}$&$0.13^{+0.06+0.07}_{-0.06-0.07}$\\
$F(q^2_{max})$&$0.75^{+0.00+0.03}_{-0.00-0.04}$&$0.59^{+0.00+0.03}_{-0.00-0.04}$&$1.38^{+0.01+0.01}_{-0.02-0.04}$&$0.45^{+0.02+0.01}_{-0.02-0.00}$&$0.54^{+0.01+0.01}_{-0.01-0.00}$&$0.11^{+0.06+0.07}_{-0.05-0.06}$\\
$a$&$0.41^{+0.01+0.05}_{-0.01-0.07}$&$-1.07^{+0.10+0.38}_{-0.10-0.32}$&$-0.23^{+0.13+0.20}_{-0.14-0.25}$&$0.30^{+0.05+0.04}_{-0.06-0.00}$&$0.41^{+0.04+0.00}_{-0.05-0.03}$&$-1.08^{+0.50+0.60}_{-0.79-0.93}$\\
$b$ &$1.40^{+0.05+0.30}_{-0.05-0.24}$&$4.85^{+0.37+1.77}_{-0.35-1.37}$&$8.74^{+0.87+1.78}_{-0.80-1.45}$&$2.07^{+0.22+0.74}_{-0.20-0.51}$&$2.16^{+0.30+0.72}_{-0.26-0.52}$&$1.98^{+0.23+0.57}_{-0.09-0.04}$\\
\hline\hline
$$&$\eta_{c}\rightarrow D_{s}$&$$& $$&$J/\psi\rightarrow D_{s} $&$$&$$\\
\hline\hline
$$&$F_{1}$& $F_{0}$&$V$&$A_{0}$&$A_{1}$&$A_{2}$\\
\hline
$F(0)$&$0.82^{+0.00+0.02}_{-0.00-0.07}$&$0.82^{+0.00+0.02}_{-0.00-0.07}$&$1.81^{+0.00+0.04}_{-0.01-0.06}$&$0.49^{+0.02+0.07}_{-0.03-0.05}$&$0.59^{+0.02+0.02}_{-0.02-0.05}$&$0.08^{+0.01+0.06}_{-0.02-0.06}$\\
$F(q^2_{max})$&$0.86^{+0.00+0.03}_{-0.00-0.07}$&$0.75^{+0.00+0.06}_{-0.00-0.07}$&$1.69^{+0.00+0.00}_{-0.01-0.02}$&$0.50^{+0.02+0.06}_{-0.03-0.06}$&$0.61^{+0.02+0.02}_{-0.02-0.05}$&$0.06^{+0.01+0.06}_{-0.02-0.05}$\\
$a$&$0.49^{+0.01+0.10}_{-0.01-0.02}$&$-0.49^{+0.07+0.44}_{-0.08-0.32}$&$0.20^{+0.08+0.26}_{-0.08-0.48}$&$0.34^{+0.03+0.05}_{-0.04-0.20}$&$0.45^{+0.02+0.13}_{-0.03-0.00}$&$-1.59^{+0.65+0.44}_{-0.87-1.51}$\\
$b$&$0.88^{+0.03+0.06}_{-0.03-0.03}$&$2.49^{+0.22+2.01}_{-0.24-1.63}$&$5.60^{+0.48+3.43}_{-0.44-2.13}$&$1.23^{+0.12+1.38}_{-0.11-0.60}$&$1.29^{+0.15+1.34}_{-0.14-0.63}$&$1.58^{+0.32+1.00}_{-0.24-0.01}$\\
 \hline\hline
 $$&$\eta_{c}(2S)\rightarrow D$&$$& $$&$\psi(2S)\rightarrow D $&$$&$$\\
\hline\hline
$$&$F_{1}$& $F_{0}$&$V$&$A_{0}$&$A_{1}$&$A_{2}$\\
\hline
$F(0)$&$0.36^{+0.02+0.03}_{-0.09-0.03}$&$0.36^{+0.02+0.03}_{-0.09-0.03}$&$0.83^{+0.16+0.07}_{-0.17-0.08}$&$0.31^{+0.04+0.02}_{-0.08-0.02}$&$0.31^{+0.00+0.00}_{-0.03-0.00}$&$0.32^{+0.16+0.09}_{-0.24-0.09}$\\
$F(q^2_{max})$&$0.37^{+0.00+0.03}_{-0.07-0.04}$&$0.39^{+0.02+0.02}_{-0.14-0.03}$&$0.61^{+0.05+0.06}_{-0.12-0.07}$&$0.28^{+0.04+0.01}_{-0.09-0.01}$&$0.30^{+0.03+0.02}_{-0.03-0.01}$&$0.19^{+0.08+0.01}_{-0.14-0.04}$\\
$a$&$0.72^{+0.57+0.07}_{-0.12-0.11}$&$0.77^{+0.21+0.03}_{-0.72-0.10}$&$0.68^{+0.10+0.14}_{-0.95-0.23}$&$0.62^{+0.31+0.01}_{-0.54-0.09}$&$0.65^{+0.09+0.01}_{-0.28-0.06}$&$-0.09^{+0.24+0.15}_{-0.11-0.33}$\\
$b$ &$2.52^{+3.22+0.26}_{-1.59-0.25}$&$1.72^{+0.98+0.55}_{-0.55-0.41}$&$8.78^{+7.33+0.46}_{-4.23-0.47}$&$4.31^{+1.27+1.76}_{-0.92-1.19}$&$3.31^{+2.18+1.07}_{-1.13-0.77}$&$10.52^{+2.47+6.01}_{-3.73-4.02}$\\
\hline\hline
$$&$\eta_{c}(2S)\rightarrow D_{s}$&$$& $$&$\psi(2S)\rightarrow D_{s} $&$$&$$\\
\hline\hline
$$&$F_{1}$& $F_{0}$&$V$&$A_{0}$&$A_{1}$&$A_{2}$\\
\hline
$F(0)$&$0.44^{+0.03+0.02}_{-0.05-0.07}$&$0.44^{+0.03+0.02}_{-0.05-0.07}$&$0.99^{+0.08+0.05}_{-0.12-0.16}$&$0.30^{+0.05+0.06}_{-0.06-0.07}$&$0.33^{+0.01+0.01}_{-0.04-0.03}$&$0.19^{+0.16+0.24}_{-0.19-0.20}$\\
$F(q^2_{max})$&$0.48^{+0.00+0.01}_{-0.07-0.08}$&$0.50^{+0.06+0.03}_{-0.06-0.10}$&$0.86^{+0.05+0.01}_{-0.06-0.13}$&$0.30^{+0.06+0.05}_{-0.07-0.08}$&$0.35^{+0.00+0.02}_{-0.04-0.03}$&$0.15^{+0.15+0.16}_{-0.15-0.15}$\\
$a$&$0.83^{+0.05+0.00}_{-0.34-0.15}$&$0.79^{+0.20+0.25}_{-0.05-0.33}$&$0.91^{+0.28+0.08}_{-0.64-0.32}$&$0.54^{+0.24+0.16}_{-0.39-0.48}$&$0.71^{+0.07+0.00}_{-0.17-0.23}$&$-0.39^{+0.79+0.48}_{-1.71-1.96}$\\
$b$&$1.87^{+1.57+0.61}_{-1.06-0.57}$&$1.00^{+1.95+1.09}_{-0.07-0.53}$&$7.45^{+4.87+1.33}_{-3.00-1.71}$&$2.32^{+0.70+1.63}_{-0.47-1.09}$&$1.99^{+1.14+1.93}_{-0.65-0.91}$&$4.21^{+1.04+3.71}_{-0.27-1.26}$\\
\hline\hline
\end{tabular}}\label{table2}
\end{center}
\end{table}
In Table \ref{form}, we compare the values of form factors at maximum recoil ($q^{2}=0$) with those obtained within the nonrelativistic quantum chromodynamics
(NRQCD) \cite{Sun:2015bxp}, the BSW model \cite{R.R} and  the QCDSR \cite{Y.M}. It is found that our predictions for the form factors of the transitions $\eta_{c}\to D_{(s)}, J/\psi \to D_{(s)}$  are comparable with those given in the NRQCD and the BSW model with the parameter $\omega=0.5$ GeV. Certainly, our results are also consistent with the previous CLFQM calculations \cite{Shen:2008zzb} within errors. While those form factors predicted in the QCDSR \cite{Y.M} are much smaller than other theoretical predictions. As to the form factors of the  $\eta_{c}(2S)\to D_{(s)}, \psi(2S)\to D_{(s)}$ transitions, only the theoretical results from the NRQCD approach are available, there exist obvious differences for some of values between these two approaches.
\begin{table}[H]
\caption{Numerical values of the transition form factors at $q^{2}= 0$, together with other theoretical results.}
\begin{center}
\begin{tabular}{|c|c|ccccc|c|}
\hline\hline
Transition  &Reference&\;\;$F_{0}(0)\;\;$&\;\;$V(0)$&\;\;$A_{0}(0)$&\;\;$A_{1}(0)$&\;\;$A_{2}(0)$\\
\hline
$\eta_{c},J/\psi\rightarrow D $&This work&$0.73$&$1.73$&$0.45$&$0.53$&$0.13$\\
\hline
$ $&\cite{Sun:2015bxp}&$0.85$&$1.76$&$0.85$&$0.72$&$-$\\
$ $&\cite{R.R}\footnotemark[1]&$-$&$2.14$&$0.55$&$0.77$&$0.31$\\
$ $&\cite{R.R}\footnotemark[2]&$-$&$2.21$&$0.54$&$0.80$&$0.47$\\
$ $&\cite{Czarnecki}\footnotemark[3]&$-$&$1.82$&$0.61$&$0.68$&$0.33$\\
$$&\cite{Shen:2008zzb}&$-$&$1.6$&$0.68$&$0.68$&$0.18$\\
$ $&\cite{Y.M}&$-$&$0.81$&$0.27$&$0.27$&$-$\\
\hline
$\eta_{c},J/\psi\rightarrow D_{s} $&This work&$0.82$&$1.81$&$0.49$&$0.59$&$0.08$\\
\hline
$ $&\cite{Sun:2015bxp}&$0.90$&$1.55$&$0.90$&$0.81$&$-$\\
$ $&\cite{R.R}\footnotemark[1]&$-$&$2.30$&$0.71$&$0.94$&$0.33$\\
$ $&\cite{R.R}\footnotemark[2]&$-$&$2.36$&$0.69$&$0.96$&$0.51$\\
$ $&\cite{Czarnecki}\footnotemark[3]&$-$&$1.80$&$0.66$&$0.78$&$0.12$\\
$$&\cite{Shen:2008zzb}&$-$&$1.8$&$0.68$&$0.68$&$0.13$\\
$ $&\cite{Y.M}&$-$&$1.07$&$0.37$&$0.38$&$-$\\
\hline
$\eta_{c}(2S),\psi(2S)\rightarrow D $&This work&$0.36$&$0.83$&$0.31$&$0.31$&$0.32$\\
\hline
$ $&\cite{Sun:2015bxp}&$0.62$&$1.00$&$0.61$&$0.54$&$-$\\
\hline
$\eta_{c}(2S),\psi(2S)\rightarrow D_{s} $&This work&$0.44$&$0.99$&$0.30$&$0.33$&$0.19$\\
\hline
$ $&\cite{Sun:2015bxp}&$0.65$&$0.83$&$0.64$&$0.59$&$-$\\
\hline\hline
\end{tabular}\label{form}
\end{center}
{\footnotesize $^1$ The form factors are computed with flavor dependent parameter $\omega$ using the WSB model.\\
               $^2$ The form factors are computed with the QCD inspired parameter $\omega = m\alpha_{s}$ using in the WSB model.\\
               $^3$ The form factors are computed with the parameter $\omega=0.5$ GeV using the WSB model.}

\end{table}

\begin{figure}[H]
\vspace{0.50cm}
  \centering
  \subfigure[]{\includegraphics[width=0.22\textwidth]{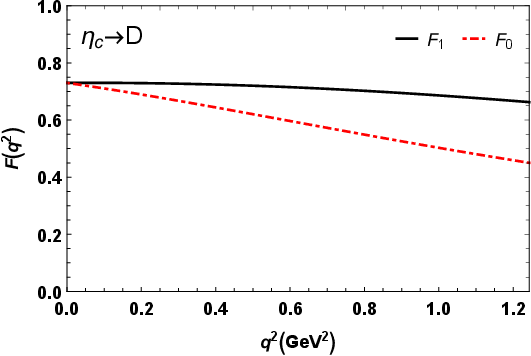}\quad}
  \subfigure[]{\includegraphics[width=0.22\textwidth]{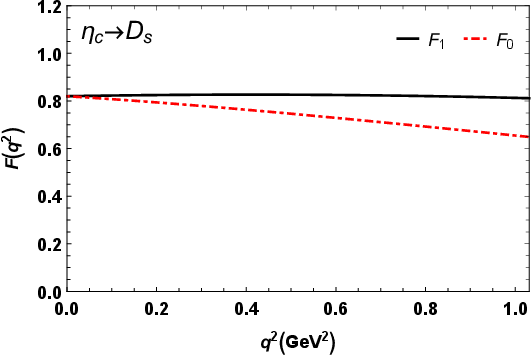}\quad}
  \subfigure[]{\includegraphics[width=0.22\textwidth]{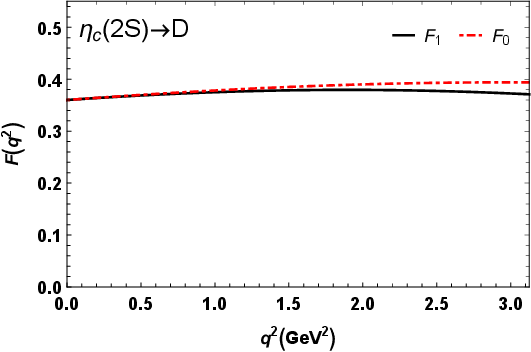}\quad}
  \subfigure[]{\includegraphics[width=0.22\textwidth]{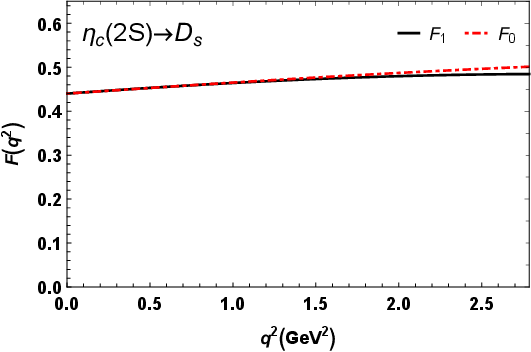}}\\
  \subfigure[]{\includegraphics[width=0.22\textwidth]{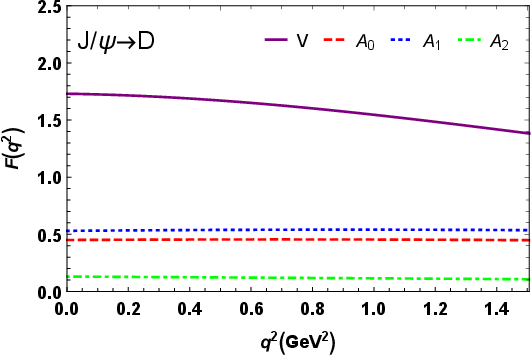}\quad}
  \subfigure[]{\includegraphics[width=0.22\textwidth]{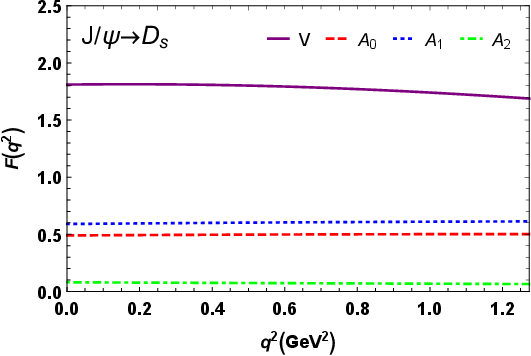}\quad}
  \subfigure[]{\includegraphics[width=0.22\textwidth]{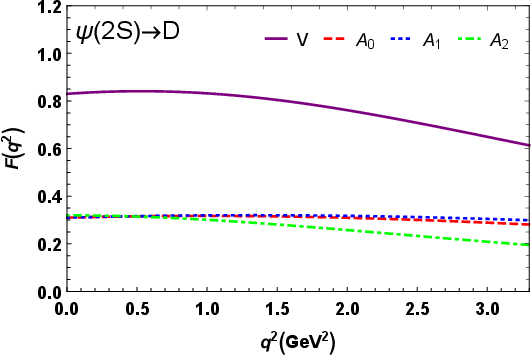}\quad}
  \subfigure[]{\includegraphics[width=0.22\textwidth]{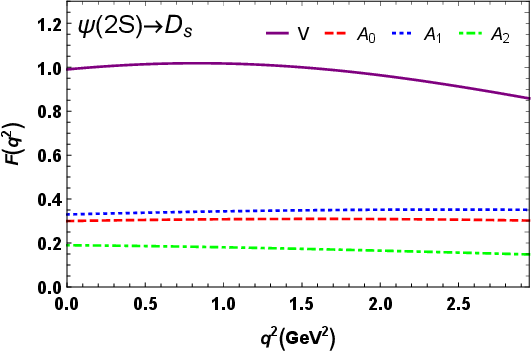}}
\caption{Form factors $F_{1}(q^2)$ and $F_{0}(q^2)$ for the transitions $\eta_{c}(1S,2S)\rightarrow D_{(s)}$  and form factors $V(q^2)$, $A_{0}(q^2)$, $A_{1}(q^2)$ and $A_{2}(q^2)$ for the transitions $\psi(1S,2S)\rightarrow D_{(s)}$, respectively.}\label{fig:T5}
\end{figure}

We plot the $q^2$-dependences of the $\eta_{c}(1S,2S)\to D_{(s)} $ and $ \psi(1S,2S)\to D_{(s)}$ transition form factors shown in Figure \ref{fig:T5}. It is very different for the $q^2$-dependences of the form factors $F_0(q^2)$ between the transitions $\eta_c(1S)\to D_{(s)}$ and $\eta_c(2S)\to D_{(s)}$. Among the form factors of the transition $J/\Psi (\psi(2S))\to D_{(s)}$, $V^{J/\Psi D_{(s)}} (V^{\psi(2S)D_{(s)}})$ is the most sensitive to the $q^2$ variation compared with other three form factors.
\subsection{Semileptonic decays}
The semileptonic decay of heavy flavor mesons offers a excellent platform for extraction of the Cabibbo-Kobayashi-Maskawa (CKM) matrix elements, which describe the CP-violating and flavor changing processes in the Standard Model. The form factors involving the dynamical information play an essential role in these semileptonic decays. Based on the form factors and the helicity amplitudes provided in the previous section, the branching ratios of the semileptonic $\eta_{c}(1S,2S)$ and $\psi(1S,2S)$ decays are presented in Table \ref{bq}, where the uncertainties arise from
the decay widths of initial charmonia, the decay
constants of initial and final state mesons, respectively. Several remarks are in order
\begin{enumerate}
\item
  For these semileptonic $\eta_{c}(1S, 2S)$ and $\psi(1S, 2S)$ decays,
   their branching ratios are in the range $10^{-14}\sim10^{-12}$ and $10^{-11}\sim10^{-10}$, respectively. Some of these decays might be detected
    by the future high-luminosity experiments, such as the Super Tau-Charm Factory (STCF), BESIII and LHC.
\item
  Our predictions for the branching ratios of the decays $J/\Psi\to D^-_{(s)}\ell^{+}\nu_{\ell}$ are consistent with those given
in the BSW model \cite{R.R}. Certainly, they are also agreement with the previous CLFQM estimates and the differences are mainly from the
input parameters. While these results are some three or more times as large as those given by the BS approach \cite{Wang:2016dkd}, the CCQM \cite{Ivanov:2015woa} and the QCDSM \cite{Wang:2007ys}. Except the variations from the input parameters, the main reason is the distinct treatment of nonperturbative dynamics, which can to be clarified by the future accurate measurements. At present BESIII only gives some upper limits, which are still much larger than all the theoretical values.
\begin{table}[H]
\caption{The branching ratios of the semileptonic $\eta_c(1S, 2S)$ and $\psi(1S, 2S)$ decays.}
\begin{center}
\scalebox{0.7}{
\begin{tabular}{|c|c|c|c|c|}
\hline\hline
$$&$10^{-14} \times \mathcal{B} r(\eta_{c}\to D^{-} e^{+}\nu_{e})$&$10^{-14} \times \mathcal{B} r(\eta_{c}\to D^{-} \mu^{+}\nu_{\mu})$&$10^{-13} \times \mathcal{B} r(\eta_{c}\to D^{-}_{s}e^{+}\nu_{e})$&$10^{-13} \times \mathcal{B} r(\eta_{c}\to D^{-}_{s}\mu^{+}\nu_{\mu})$\\
\hline
This work&$5.43^{+0.12+0.00+0.47}_{-0.12-0.00-0.60}$&$5.15^{+0.12+0.00+0.45}_{-0.11-0.00-0.57}$&$8.97^{+0.20+0.01+0.51}_{-0.19-0.01-1.48}$&$8.46^{+0.19+0.01+0.50}_{-0.18-0.01-1.40}$\\
\hline\hline
$$&$10^{-11} \times \mathcal{B} r(J/\psi \to D^{-} e^{+}\nu_{e})$&$10^{-11} \times \mathcal{B} r(J/\psi\to D^{-} \mu^{+}\nu_{\mu})$&$10^{-10} \times \mathcal{B} r(J/\psi\to D^{-}_{s}e^{+}\nu_{e})$&$10^{-10} \times \mathcal{B} r(J/\psi\to D^{-}_{s}\mu^{+}\nu_{\mu})$\\
\hline
This work&$6.10^{+0.11+0.10+0.14}_{-0.11-0.12-0.19}$&$5.78^{+0.11+0.11+0.16}_{-0.10-0.13-0.11}$&$10.21^{+0.19+0.66+0.56}_{-0.18-0.61-1.41}$&$9.59^{+0.18+0.62+0.63}_{-0.17-0.58-1.34}$\\
\hline
QCDSR\cite{Wang:2007ys}&$0.73$&$0.71$&$1.8$&$1.7$\\
\hline
LFQM\cite{Shen:2008zzb}&$5.1\sim5.7$&$4.7\sim5.5$&$5.3\sim5.8$&$5.5\sim5.7$\\
\hline
BSW\cite{R.R}&$6.0$&$5.8$&$10.4$&$9.93$\\
\hline
CCQM\cite{Ivanov:2015woa}&$1.71$&$1.66$&$3.3$&$3.2$\\
\hline
BS\cite{Wang:2016dkd}&$2.03$&$1.98$&$3.67$&$3.54$\\
\hline
Exp.\cite{Wang:2022ghh,BESIII:2021mnd,BESIII:2023fqz}&$<7.1\times10^{3}$&$<5.6\times10^{4}$&$<1.3\times10^{4}$&$-$\\
\hline\hline
$$&$10^{-13} \times \mathcal{B} r(\eta_{c}(2S)\to D^{-} e^{+}\nu_{e})$&$10^{-13} \times \mathcal{B} r(\eta_{c}(2S)\to D^{-} \mu^{+}\nu_{\mu})$&$10^{-12} \times \mathcal{B} r(\eta_{c}(2S)\to D^{-}_{s}e^{+}\nu_{e})$&$10^{-12} \times \mathcal{B} r(\eta_{c}(2S)\to D^{-}_{s}\mu^{+}\nu_{\mu})$\\
\hline
This work&$3.12^{+1.08+0.47+0.56}_{-0.69-1.35-0.53}$&$3.08^{+1.06+0.47+0.56}_{-0.68-1.34-0.52}$&$7.36^{+2.54+0.93+0.62}_{-1.62-1.75-2.22}$&$7.25^{+2.50+0.91+0.60}_{-1.60-1.73-2.19}$\\
\hline\hline
$$&$10^{-11} \times \mathcal{B} r(\psi(2S) \to D^{-} e^{+}\nu_{e})$&$10^{-11} \times \mathcal{B} r(\psi(2S)\to D^{-} \mu^{+}\nu_{\mu})$&$10^{-10} \times \mathcal{B} r(\psi(2S)\to D^{-}_{s}e^{+}\nu_{e})$&$10^{-10} \times \mathcal{B} r(\psi(2S)\to D^{-}_{s}\mu^{+}\nu_{\mu})$\\
\hline
This work&$3.45^{+0.10+0.49+0.23}_{-0.09-0.20-0.25}$&$3.39^{+0.09+0.11+0.21}_{-0.09-0.35-0.23}$&$7.20^{+0.20+0.97+0.60}_{-0.19-0.44-0.92}$&$7.02^{+0.20+0.99+0.65}_{-0.19-0.38-0.83}$\\
\hline\hline
\end{tabular}\label{bq}}
\end{center}
\end{table}
\item The branching ratios of the decays $\eta_c(2S)\to D^-_{(s)}\ell^+\nu_{\ell}$ are larger than those of the decays $\eta_c\to D^-_{(s)}\ell^+\nu_{\ell}$. It is contrary for the cases of the decays $\psi(1S,2S)\to D^-_{(s)}\ell^+\nu_{\ell}$ , where $Br(\psi(2S)\to D^-_{(s)}\ell^+\nu_{\ell})<Br(J/\Psi\to D^-_{(s)}\ell^+\nu_{\ell})$. These are related with their total widths, $\Gamma_{\eta_c}(\Gamma_{\psi(2S)})$ is about 3 times as large as $\Gamma_{\eta_c(2S)}(\Gamma_{J/\Psi})$.

\item
 In order to cancel out a large part of the theoretical and experimental uncertainties, to check the lepton flavor universality (LFU) and to detect the effect of SU(3) symmetry breaking,  it is helpful to consider the ratio $R\equiv\mathcal{B} r(\eta_c(\psi)\to D^{-}_{s}\ell^{+}\nu_{\ell})/\mathcal{B} r(\eta_c(\psi)\to D^{-}\ell^{+}\nu_{\ell})$, which should be equal to $|V_{cs}/V_{cd}|^2\approx19.46$ under the $SU(3)$ flavor symmetry limit. Their values in this work are listed as
 \begin{footnotesize}
\begin{equation}
\begin{aligned}
R^{e}_{\eta_{c}}&=&\frac{\eta_{c}\to D^{-}_{s}e^{+}\nu_{e}}{\eta_{c}\to D^{-}e^{+}\nu_{e}}=16.52\pm2.73,\;\;\;
R^{e}_{\eta_{c}(2S)}=\frac{\eta_{c}(2S)\to D^{-}_{s}e^{+}\nu_{e}}{\eta_{c}(2S)\to D^{-}e^{+}\nu_{e}}=23.59\pm13.06 ,\\
R^{\mu}_{\eta_{c}}&=&\frac{\eta_{c}\to D^{-}_{s}\mu^{+}\nu_{\mu}}{\eta_{c}\to D^{-}\mu^{+}\nu_{\mu}}=16.43\pm3.27,\;\;\;
R^{\mu}_{\eta_{c}(2S)}=\frac{\eta_{c}(2S)\to D^{-}_{s}\mu^{+}\nu_{\mu}}{\eta_{c}(2S)\to D^{-}\mu^{+}\nu_{\mu}}=23.54\pm13.07 ,\\
R^{e}_{J/\psi}&=&\frac{J/\psi\to D^{-}_{s}e^{+}\nu_{e}}{J/\psi\to D^{-}e^{+}\nu_{e}}=16.74\pm2.37, \;\;\;\;
R^{e}_{\psi(2S)}=\frac{\psi(2S)\to D^{-}_{s}e^{+}\nu_{e}}{\psi(2S)\to D^{-}e^{+}\nu_{e}}=20.87\pm4.09 ,\\
R^{\mu}_{J/\psi}&=&\frac{J/\psi\to D^{-}_{s}\mu^{+}\nu_{\mu}}{J/\psi\to D^{-}\mu^{+}\nu_{\mu}}=16.59\pm2.36, \;\;\;\;
R^{\mu}_{\psi(2S)}=\frac{\psi(2S)\to D^{-}_{s}\mu^{+}\nu_{\mu}}{\psi(2S)\to D^{-}\mu^{+}\nu_{\mu}}=20.71\pm3.62.
\end{aligned}
\end{equation}
\end{footnotesize}
It is obviously there exist some effects of $SU(3)$ symmetry breaking in these semi-leptonic decays. The ratios $R^{e, \mu}_{J/\psi}$ are consistent with that given in Ref. \cite{Wang:2016dkd}, where $R^{\ell}_{J/\psi}=18.1$. Certainly, the values of these ratios are in agreement with the predictions under the $SU(3)$ flavor symmetry limit within errors. The large uncertainties from the ratios $R^{\ell}_{\eta_{c}(2S)}$ are mainly induced by the decay width of $\eta_c(2S)$, $\Gamma_{\eta_c(2S)}=(11.3^{+3.2}_{-2.9})$ MeV.
\end{enumerate}
\subsection{Physical observables}
\begin{table}[H]
\caption{The forward-backward asymmetry $A_{FB}$.}
\begin{center}
\scalebox{0.8}{
\begin{tabular}{|c|c|c|c|c|}
\hline\hline
 Channel  &$\eta_{c}\to D^{-}e^{+}\nu_{e}$&$\eta_{c}\to D^{-}\mu^{+}\nu_{\mu}$&$\eta_{c}\to D^{-}_{s}e^{+}\nu_{e}$&$\eta_{c}\to D^{-}_{s}\mu^{+}\nu_{\mu}$\\
 \hline
 $A_{FB}$&$(4.21^{+0.09+0.00+0.37}_{-0.09-0.00-0.46})\times10^{-6}$&$0.080^{+0.002+0.000+0.007}_{-0.002-0.000-0.009}$&$(5.08^{+0.11+0.00+0.28}_{-0.11-0.00-0.84})\times10^{-6}$&$0.091^{+0.002+0.000+0.006}_{-0.002-0.000-0.015}$\\
 \hline
  Channel  &$J/\psi \to D^{-}e^{+}\nu_{e}$&$J/\psi\to D^{-}\mu^{+}\nu_{\mu}$&$J/\psi\to D^{-}_{s}e^{+}\nu_{e}$&$J/\psi\to D^{-}_{s}\mu^{+}\nu_{\mu}$\\
   \hline
 $ A_{FB}$&$-0.23^{+0.00+0.01+0.01}_{-0.00-0.00-0.00}$&$-0.23^{+0.00+0.01+0.01}_{-0.00-0.01-0.00}$&$-0.21^{+0.00+0.01+0.03}_{-0.00-0.01-0.01}$&$-0.22^{+0.00+0.01+0.03}_{-0.00-0.01-0.01}$\\
\hline\hline
Channel&$\eta_{c}(2S)\to D^{-}e^{+}\nu_{e}$&$\eta_{c}(2S)\to D^{-}\mu^{+}\nu_{\mu}$&$\eta_{c}(2S)\to D^{-}_{s}e^{+}\nu_{e}$&$\eta_{c}(2S)\to D^{-}_{s}\mu^{+}\nu_{\mu}$\\
\hline
$A_{FB}$&$(1.69^{+0.58+0.21+0.30}_{-0.37-0.76-0.28})\times10^{-6}$&$0.045^{+0.016+0.006+0.008}_{-0.010-0.021-0.007}$&$(1.86^{+0.64+0.27+0.19}_{-0.41-0.41-0.56})\times10^{-6}$&$0.048^{+0.017+0.007+0.005}_{-0.011-0.011-0.015}$\\
\hline
Channel&$\psi(2S) \to D^{-}e^{+}\nu_{e}$&$\psi(2S)\to D^{-}\mu^{+}\nu_{\mu}$&$\psi(2S)\to D^{-}_{s}e^{+}\nu_{e}$&$\psi(2S)\to D^{-}_{s}\mu^{+}\nu_{\mu}$\\
\hline
$A_{FB}$&$-0.28^{+0.01+0.09+0.03}_{-0.01-0.02-0.02}$&$-0.28^{+0.01+0.09+0.03}_{-0.01-0.02-0.02}$&$-0.27^{+0.01+0.07+0.07}_{-0.01-0.02-0.01}$&$-0.27^{+0.01+0.07+0.07}_{-0.01-0.02-0.01}$\\
\hline\hline
\end{tabular}\label{AFB}}
\end{center}
\end{table}
In order to study the impact of lepton mass and provide a more detailed physical picture for the semileptonic decays, we also define other two physical observables on the basis of form factors and helicity formalism, that is the forward-backward asymmetry $A_{FB}$ and the longitudinal polarization fraction $f_{L}$. The results of these two physical observables are listed in Tables \ref{AFB} and \ref{FL}, respectively. We find that the ratios of the forward-backward asymmetries $A^\mu_{FB}/ A^e_{FB}$ between the semileptonic decays $\eta_{c}\to D^-_{(s)}\mu^{+}\nu_{\mu}$ and $\eta_{c}\to D^-_{(s)}e^{+}\nu_{e}$  are about $1.9 (1.8)\times10^{4}$ for $\eta_c(1S)$ and
$2.7 (2.6)\times10^{4}$ for $\eta_c(2S)$, respectively. The reason is that the forward-backward asymmetries $A_{FB}$ for the decays $\eta_{c}(1S, 2S)\to D^-_{(s)}\ell^+\nu_{\ell}$  are proportional to the square of the lepton mass. Undoubtedly, the effect of lepton mass can be well checked in such decay mode with a pseudoscalar meson involved in the final states. It is similar to the decays $B_c\to \eta_c(1S, 2S, 3S)\ell^+\nu_{\ell}$ \cite{sunzj}. While for the decays $\psi(1S,2S) \to D^-_{(s)}\ell^+\nu_{\ell}$, the values of the forward-backward asymmetries $A^\mu_{FB}$ and $A^e_{FB}$ are almost equal to each other. It is noted that the dominant contributions to the $A_{FB}$ for the transitions $\psi(1S, 2S)\to D_{(s)}$  arise from the terms proportional to $(H^{2}_{V,+}-H^{2}_{V,-})$ in Eq. (\ref{eq:btheta2}).

In Table \ref{FL}, we can clearly find that the longitudinal polarization fractions $f_{L}$ between the decays $\psi(1S,2S) \to D^{-}_{(s)}e^{+}\nu_{e}$
and $\psi(1S,2S) \to D^{-}_{(s)}\mu^{+}\nu_{\mu}$ are very close to each other
\be
f_{L}(\psi(1S,2S) \to D^{-}_{(s)}e^{+}\nu_{e})\sim f_{L}(\psi(1S,2S) \to D^{-}_{(s)}\mu^{+}\nu_{\mu}),
\en
which reflects the lepton flavor universality (LFU).
In order to investigate the dependences of the polarizations on the different $q^2$, we calculate the longitudinal polarization fractions
by dividing the full energy region into two regions for each decay.  Region 1 is defined as $m_{\ell}^{2}<q^{2}<\frac{(m_{\psi(nS)}-m_{D_{(s)}})^{2}+m_{\ell}^{2}}{2} $ and Region 2 is $\frac{(m_{\psi(nS)}-m_{D_{(s)}})^{2}+m_{\ell}^{2}}{2} <q^{2}<(m_{\psi(nS)}-m_{D_{(s)}})^{2}$ with $n=1,2$. Interestingly, for the decays $\psi(1S, 2S) \to D^{-}_{(s)}\ell^{+}\nu_{\ell}$ the longitudinal (transverse) polarization is dominant in Region 1 (Region 2). While these two kinds of polarizations are comparable in the entire physical region. These results can be tested by the future high-luminosity experiments.

\begin{table}[H]
\caption{The partial branching ratios and the longitudinal polarization fractions $f_{L}$ for the decays $\psi(1S, 2S) \to D^{-}_{(s)}\ell^{+}\nu_{\ell}$ in Region 1 and Region 2.}
\begin{center}
\scalebox{0.7}{
\begin{tabular}{|c|c|c|c|c|c|c|c|}
\hline\hline
Observables&Region 1&Region 2&Total&Observables&Region 1&Region 2&Total\\
\hline\hline
$\mathcal{B} r(J/\psi \to D^{-}e^{+}\nu_{e})$&$3.54\times10^{-11}$&$2.55\times10^{-11}$&$6.10\times10^{-11}$&$\mathcal{B} r(J/\psi\to D^{-}\mu^{+}\nu_{\mu})$&$3.31\times10^{-11}$&$2.46\times10^{-11}$&$5.78\times10^{-11}$\\
$f_{L}(J/\psi \to D^{-}e^{+}\nu_{e})$&$0.68$&$0.42$&$0.57^{+0.01+0.00+0.03}_{-0.01-0.00-0.03}$&$f_{L}(J/\psi\to D^{-}\mu^{+}\nu_{\mu})$&$0.67$&$0.42$&$0.56^{+0.01+0.00+0.03}_{-0.01-0.01-0.02}$\\
\hline
$\mathcal{B} r(J/\psi \to D^{-}_{s}e^{+}\nu_{e})$&$5.83\times10^{-10}$&$4.38\times10^{-10}$&$10.21\times10^{-10}$&$\mathcal{B} r(J/\psi\to D^{-}_{s}\mu^{+}\nu_{\mu})$&$5.39\times10^{-10}$&$4.20\times10^{-10}$&$9.59\times10^{-10}$\\
$f_{L}(J/\psi \to D^{-}_{s}e^{+}\nu_{e})$&$0.70$&$0.43$&$0.58^{+0.01+0.04+0.03}_{-0.01-0.03-0.08}$&$f_{L}(J/\psi\to D^{-}_{s}\mu^{+}\nu_{\mu})$&$0.68$&$0.43$&$0.57^{+0.01+0.04+0.03}_{-0.01-0.03-0.08}$\\
\hline\hline
Observables&Region 1&Region 2&Total&Observables&Region 1&Region 2&Total\\
\hline\hline
$\mathcal{B} r(\psi(2S) \to D^{-}e^{+}\nu_{e})$&$1.91\times10^{-11}$&$1.54\times10^{-11}$&$3.45\times10^{-11}$&$\mathcal{B} r(\psi(2S)\to D^{-}\mu^{+}\nu_{\mu})$&$1.88\times10^{-11}$&$1.51\times10^{-11}$&$3.39\times10^{-11}$\\
$f_{L}(\psi(2S) \to D^{-}e^{+}\nu_{e})$&$0.60$&$0.40$&$0.51^{+0.01+0.07+0.08}_{-0.01-0.14-0.08}$&$f_{L}(\psi(2S)\to D^{-}\mu^{+}\nu_{\mu})$&$0.60$&$0.40$&$0.51^{+0.01+0.05+0.08}_{-0.01-0.13-0.07}$\\
\hline
$\mathcal{B} r(\psi(2S) \to D^{-}_{s}e^{+}\nu_{e})$&$4.06\times10^{-10}$&$3.14\times10^{-10}$&$7.20\times10^{-10}$&$\mathcal{B} r(\psi(2S)\to D^{-}_{s}\mu^{+}\nu_{\mu})$&$3.94\times10^{-10}$&$3.08\times10^{-10}$&$7.02\times10^{-10}$\\
$f_{L}(\psi(2S) \to D^{-}_{s}e^{+}\nu_{e})$&$0.65$&$0.41$&$0.54^{+0.02+0.02+0.02}_{-0.01-0.08-0.14}$&$f_{L}(\psi(2S)\to D^{-}_{s}\mu^{+}\nu_{\mu})$&$0.64$&$0.41$&$0.54^{+0.02+0.03+0.01}_{-0.01-0.08-0.13}$\\
\hline\hline
\end{tabular}\label{FL}}
\end{center}
\end{table}
\begin{figure}[H]
\vspace{0.40cm}
  \centering
  \subfigure[]{\includegraphics[width=0.22\textwidth]{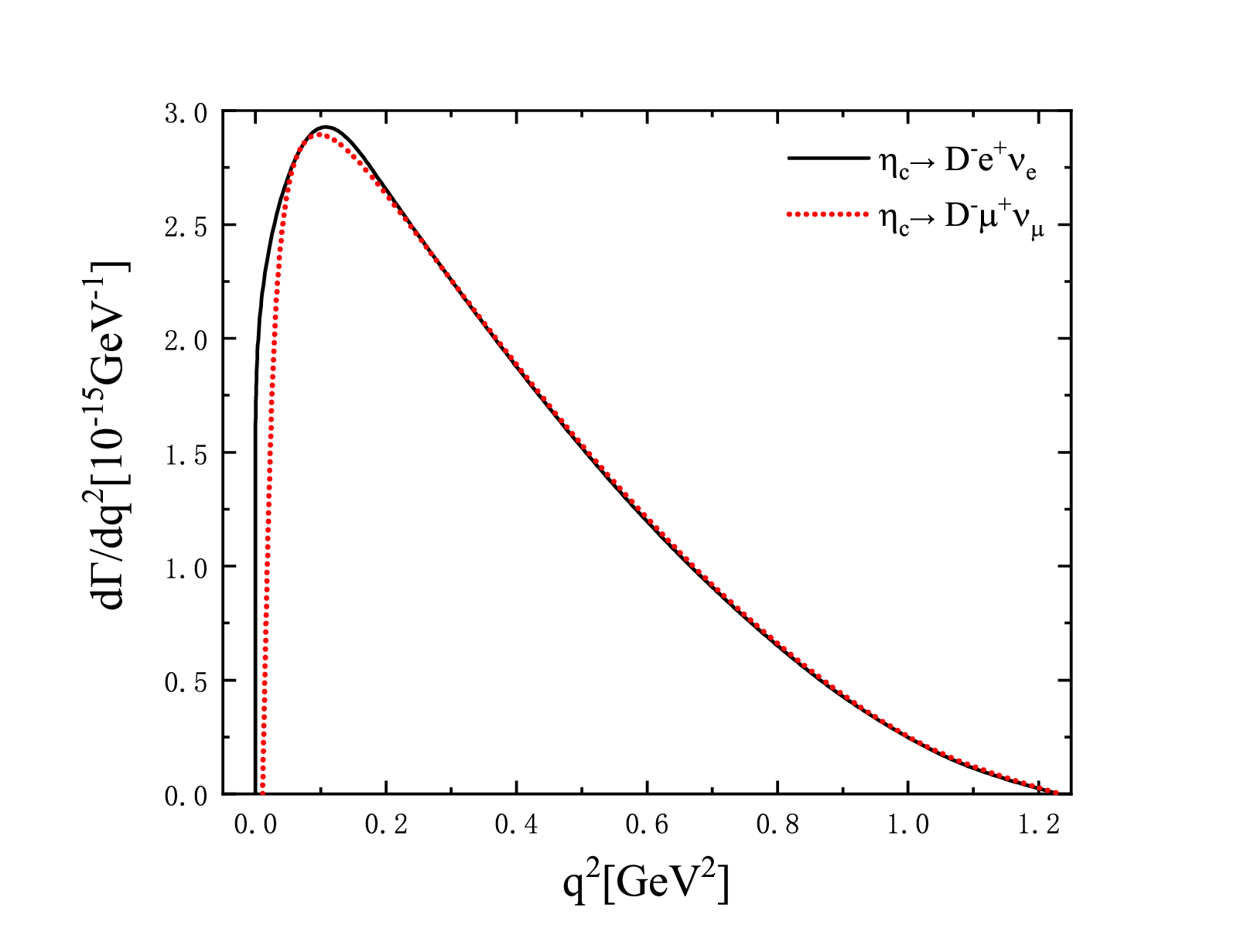}\quad}
  \subfigure[]{\includegraphics[width=0.22\textwidth]{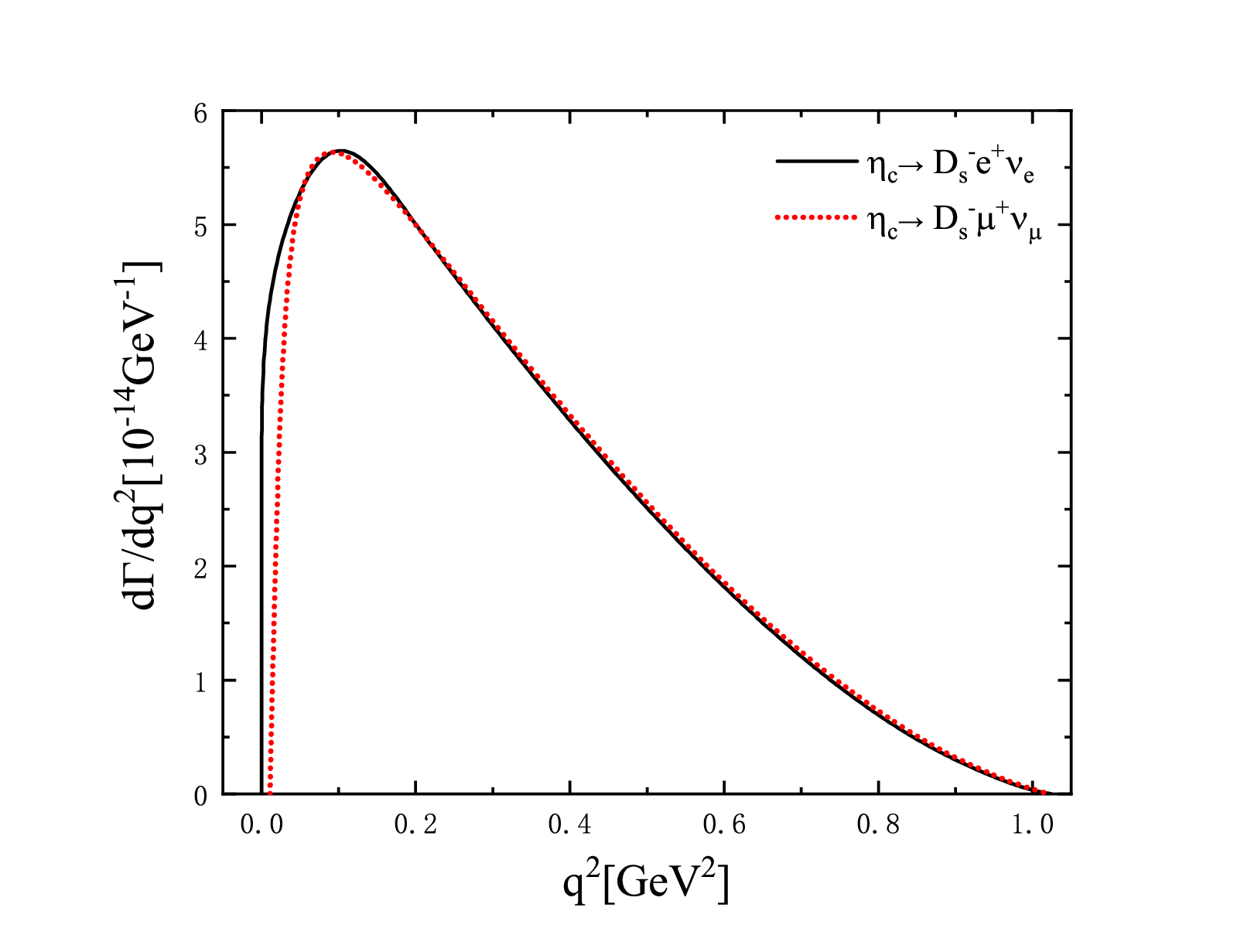}\quad}
  \subfigure[]{\includegraphics[width=0.22\textwidth]{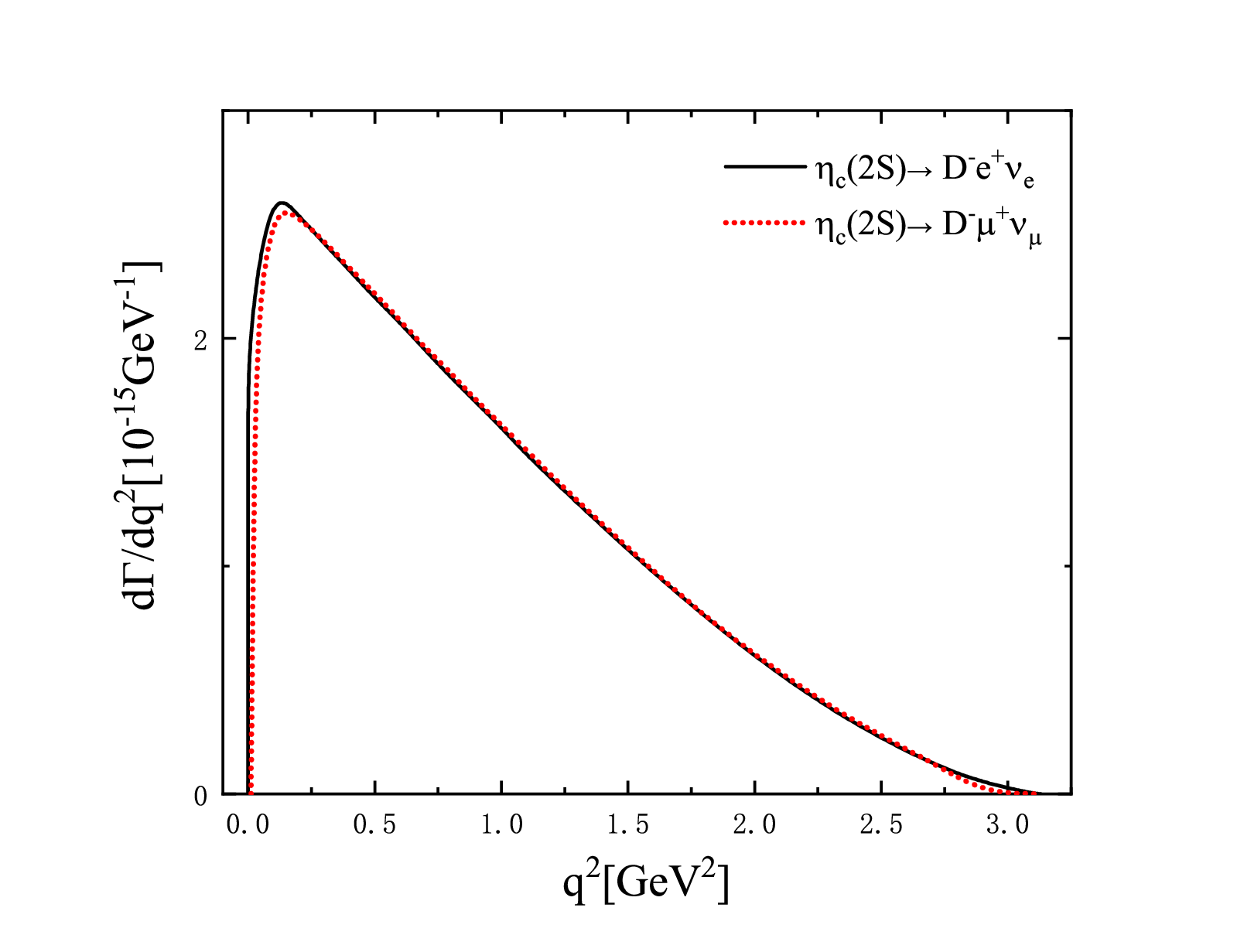}\quad}
  \subfigure[]{\includegraphics[width=0.22\textwidth]{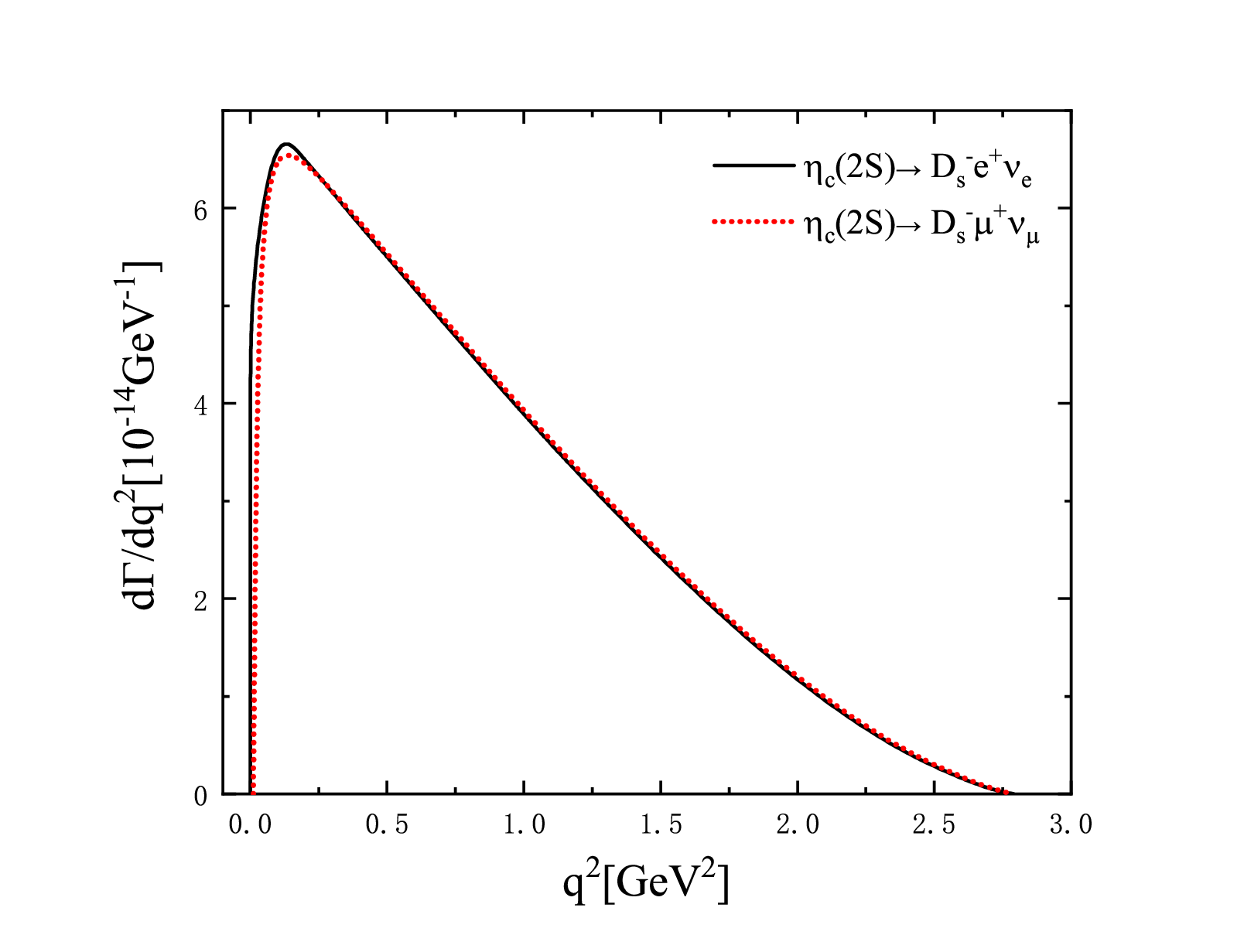}}\\
  \subfigure[]{\includegraphics[width=0.22\textwidth]{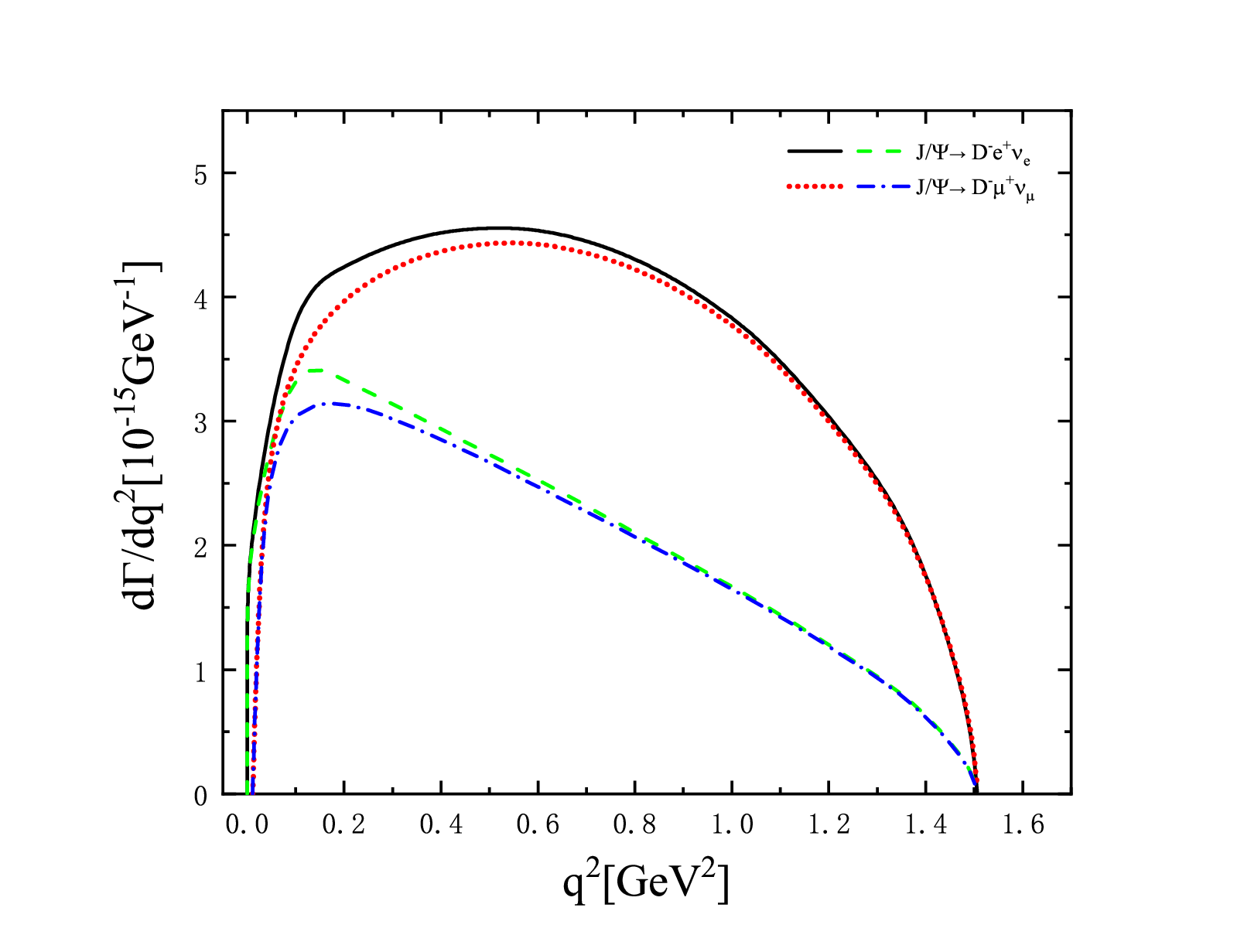}\quad}
  \subfigure[]{\includegraphics[width=0.22\textwidth]{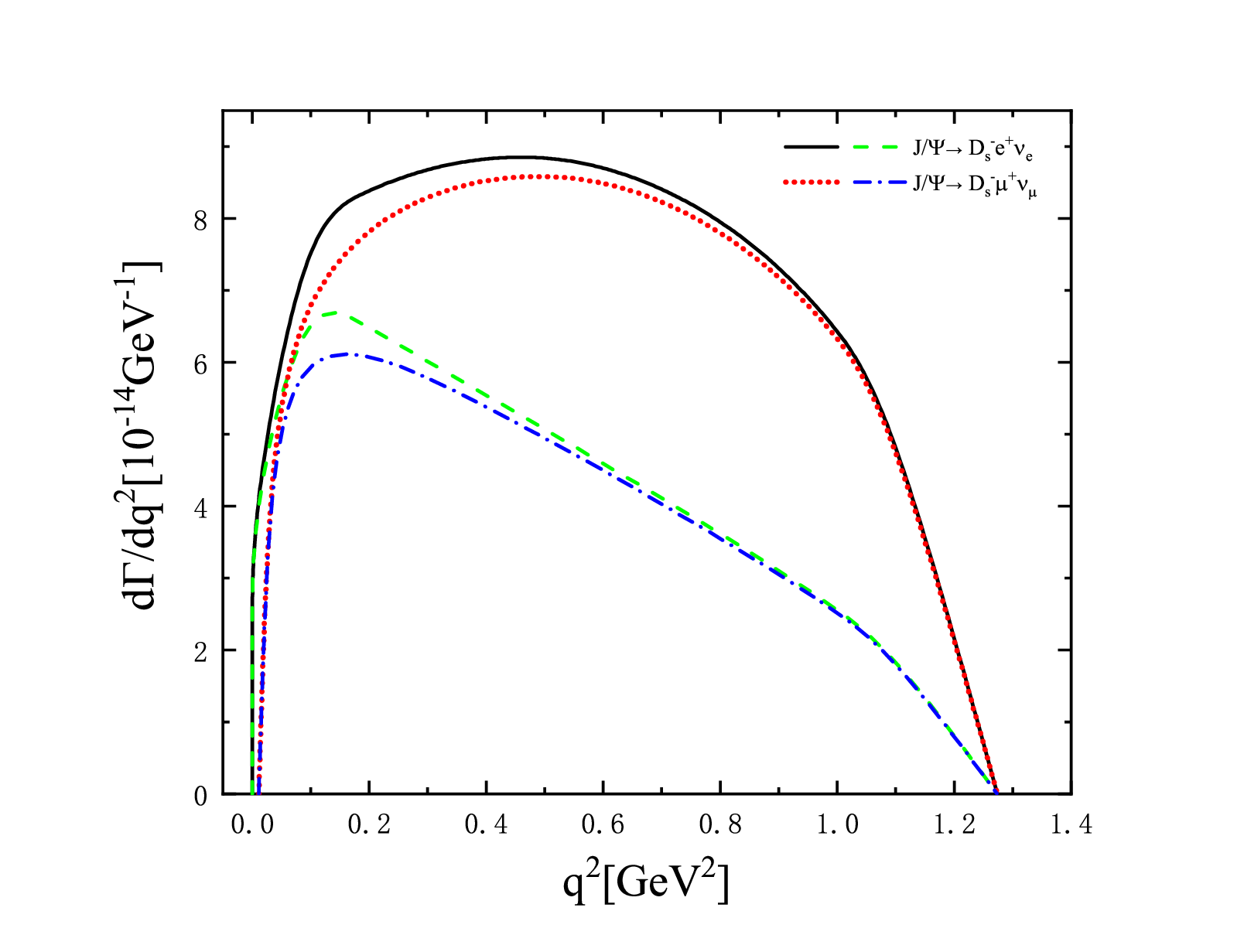}\quad}
  \subfigure[]{\includegraphics[width=0.22\textwidth]{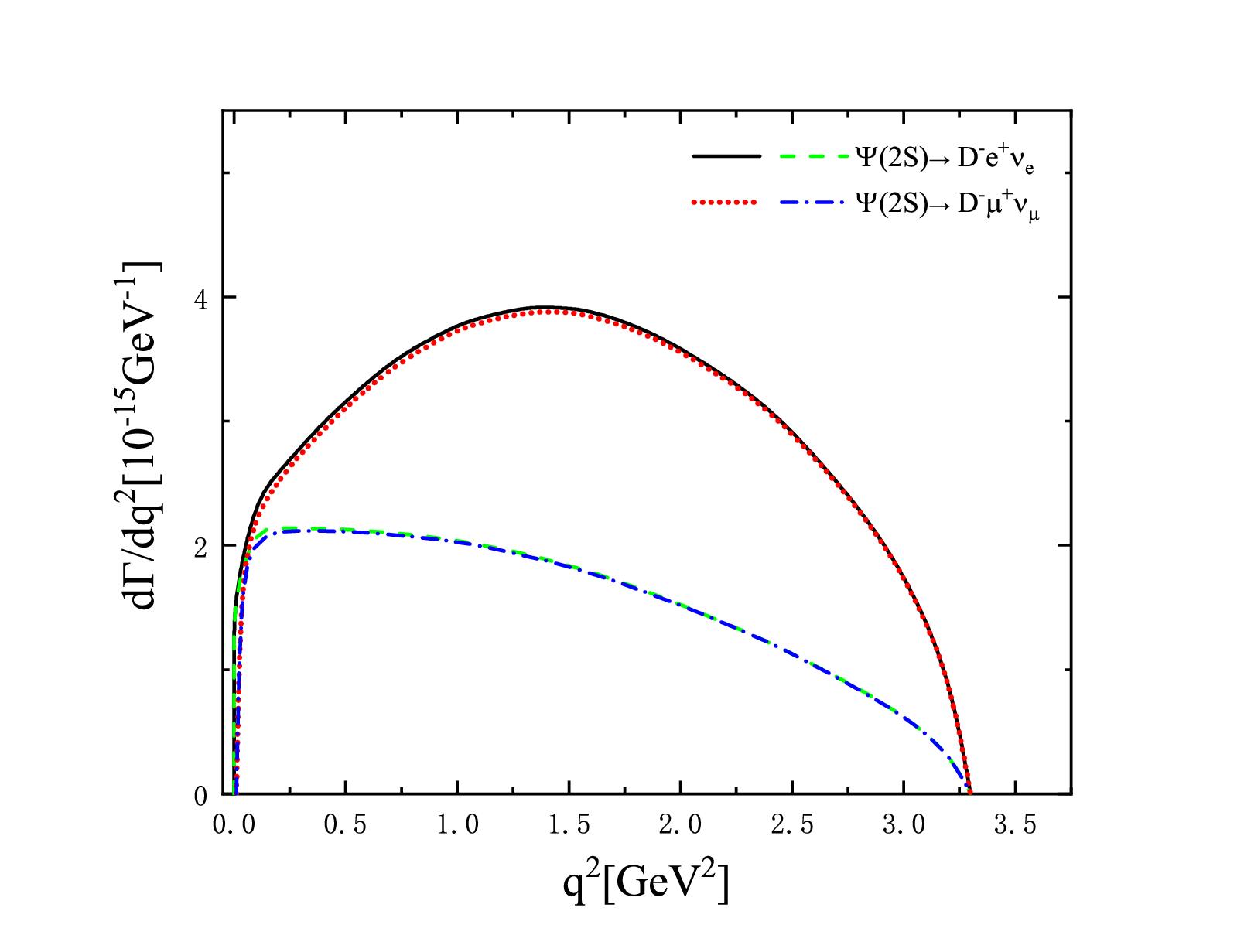}\quad}
  \subfigure[]{\includegraphics[width=0.22\textwidth]{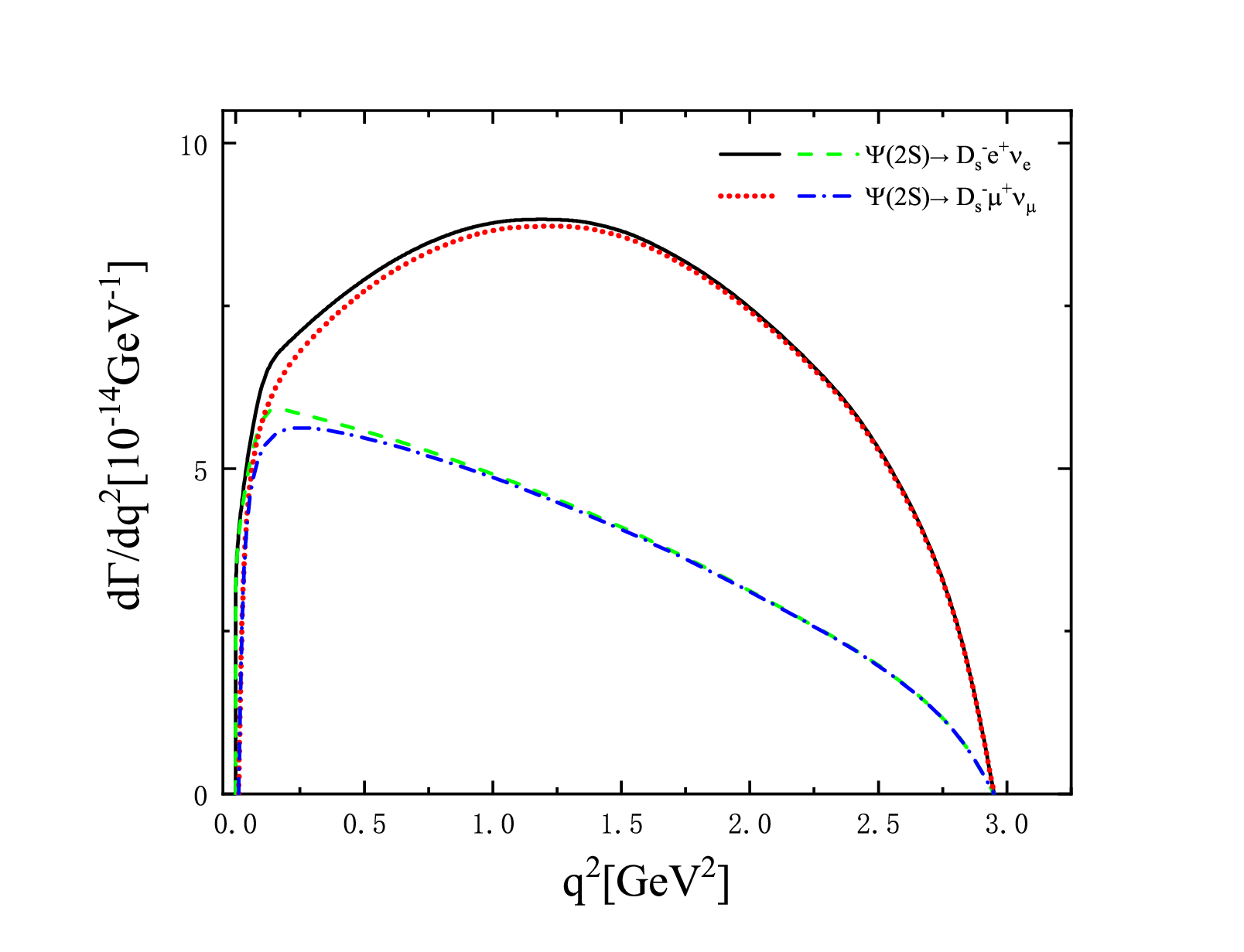}}
    \caption{The theoretical predictions for the $q^2$ dependences of the differential decay rates $d\Gamma/dq^2$ and $d\Gamma^{L}/dq^2$ .}\label{fig:T2}
\end{figure}
In Figures \ref{fig:T2}-\ref{fig:T4}, we also display the  $q^{2}$-dependences of differential decay rates $d\Gamma_{(L)}/dq^{2}$ and forward-backward asymmetries $A_{FB}$, respectively. It can be observed that the values of $d\Gamma_{(L)}/dq^{2}$ and $A_{FB}$ coincide with 0 at the zero recoil point $(q^{2}=q^{2}_{max})$ since the coefficient $\sqrt{\lambda(q^2)}=\sqrt{\lambda(m^{2}_{\eta_{c},J/\psi},m^{2}_{D_{s}},q^{2})}$ shown in Eqs.(\ref{eq:pp}-\ref{eq:btheta2}) at the same zero recoil point being equal to 0. The lepton mass effects can be obviously observed from Figure 4(a)-4(d).
\begin{figure}[H]
\vspace{0.40cm}
  \centering
  \subfigure[]{\includegraphics[width=0.22\textwidth]{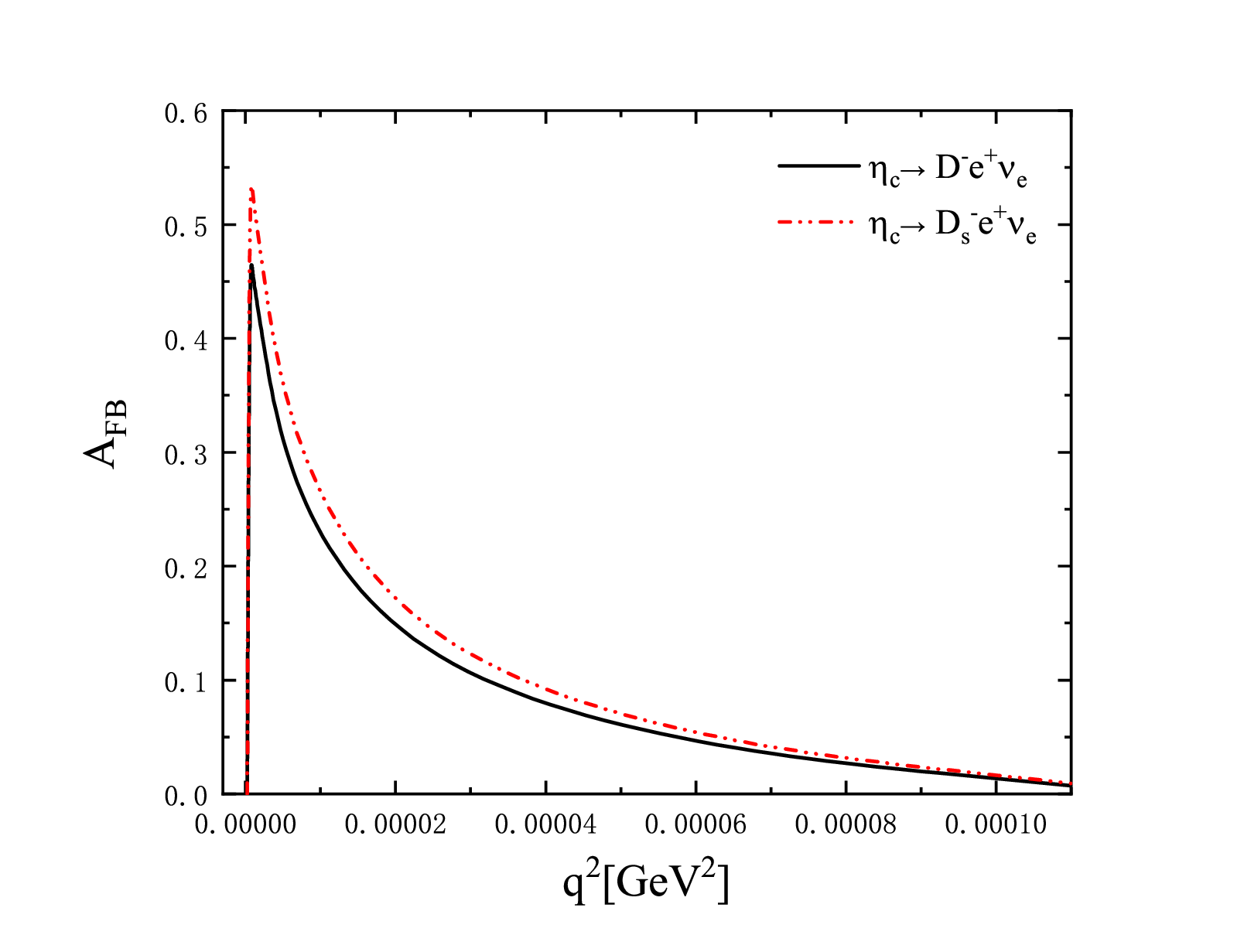}\quad}
  \subfigure[]{\includegraphics[width=0.22\textwidth]{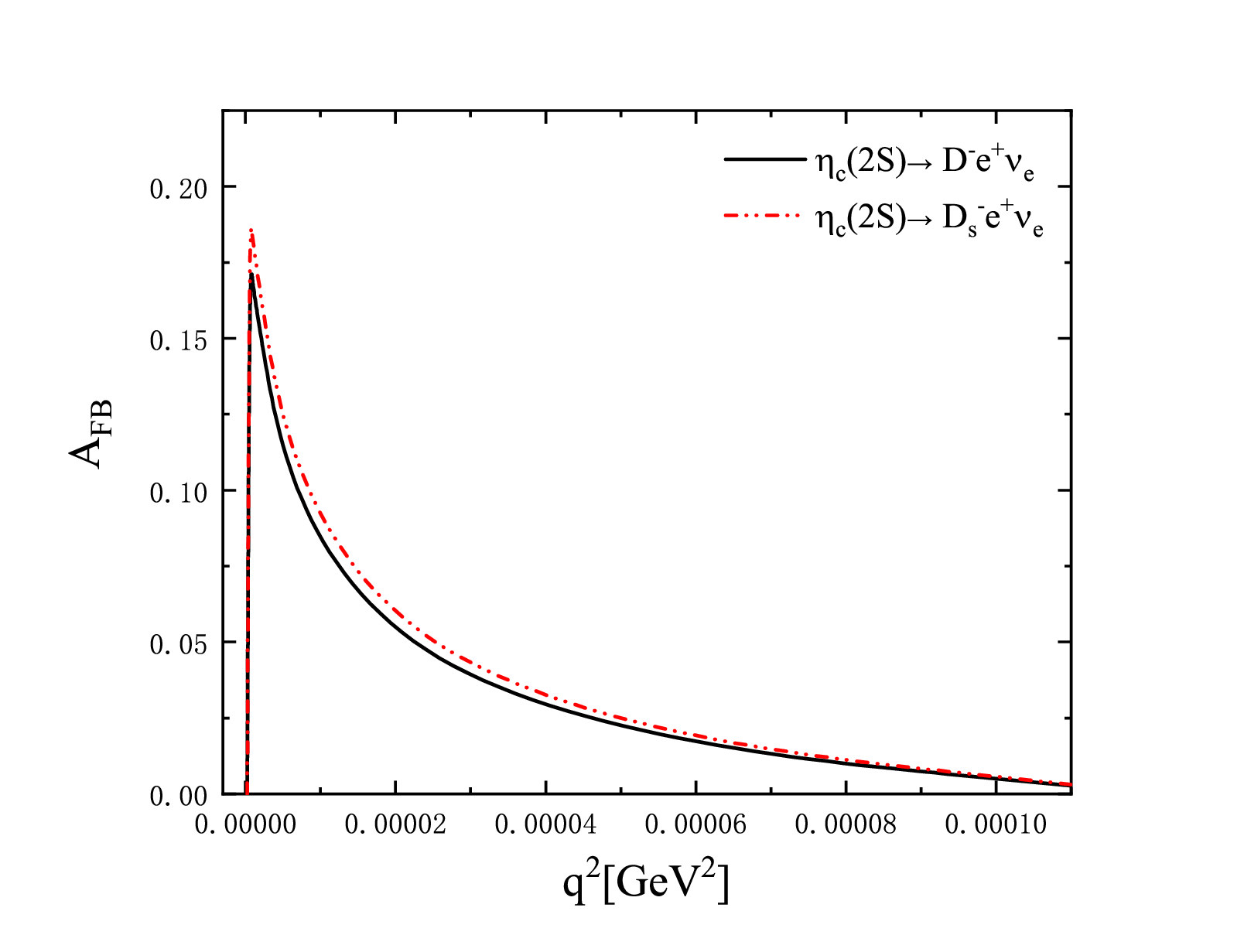}\quad}
  \subfigure[]{\includegraphics[width=0.22\textwidth]{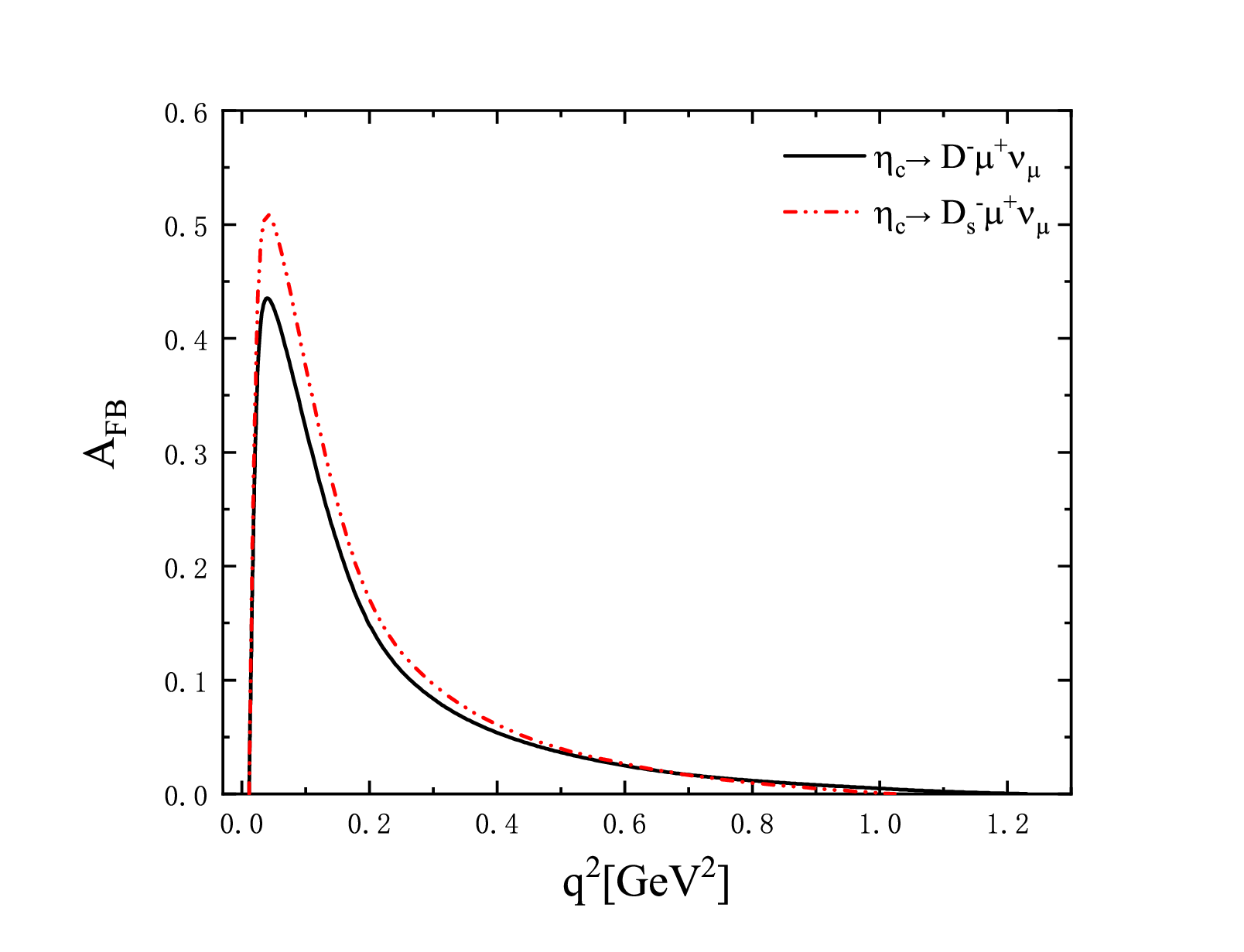}\quad}
  \subfigure[]{\includegraphics[width=0.22\textwidth]{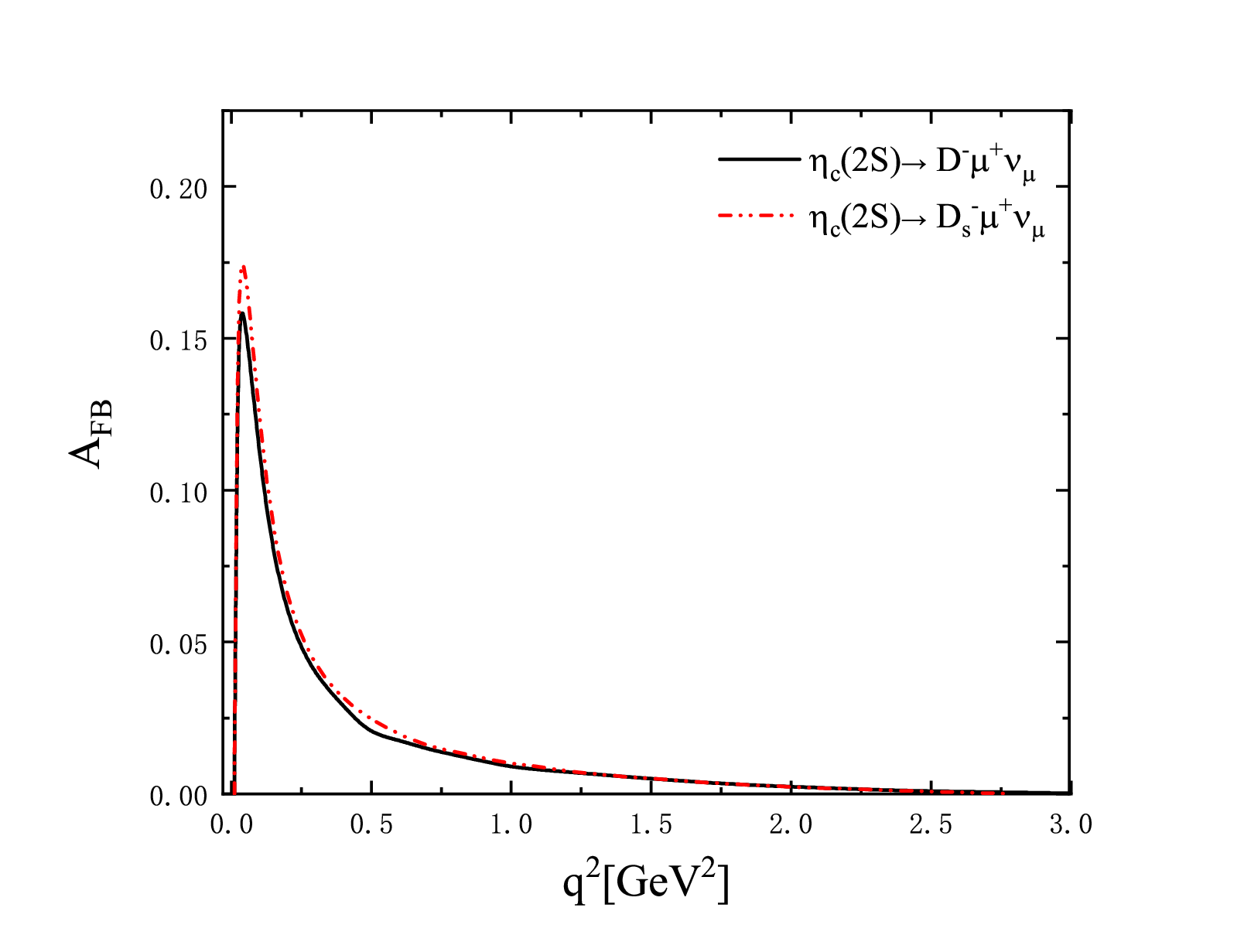}}\\
  \subfigure[]{\includegraphics[width=0.22\textwidth]{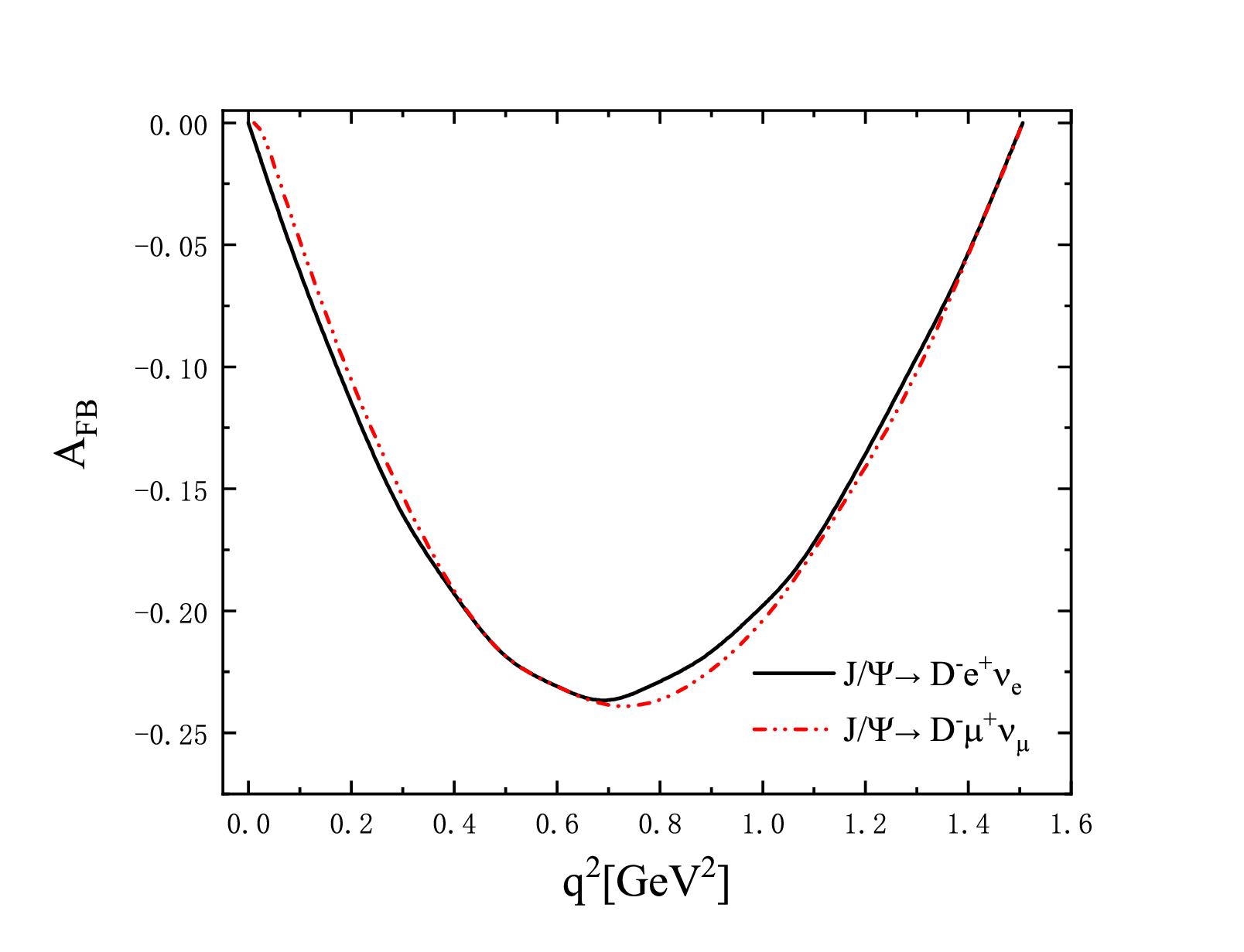}\quad}
  \subfigure[]{\includegraphics[width=0.22\textwidth]{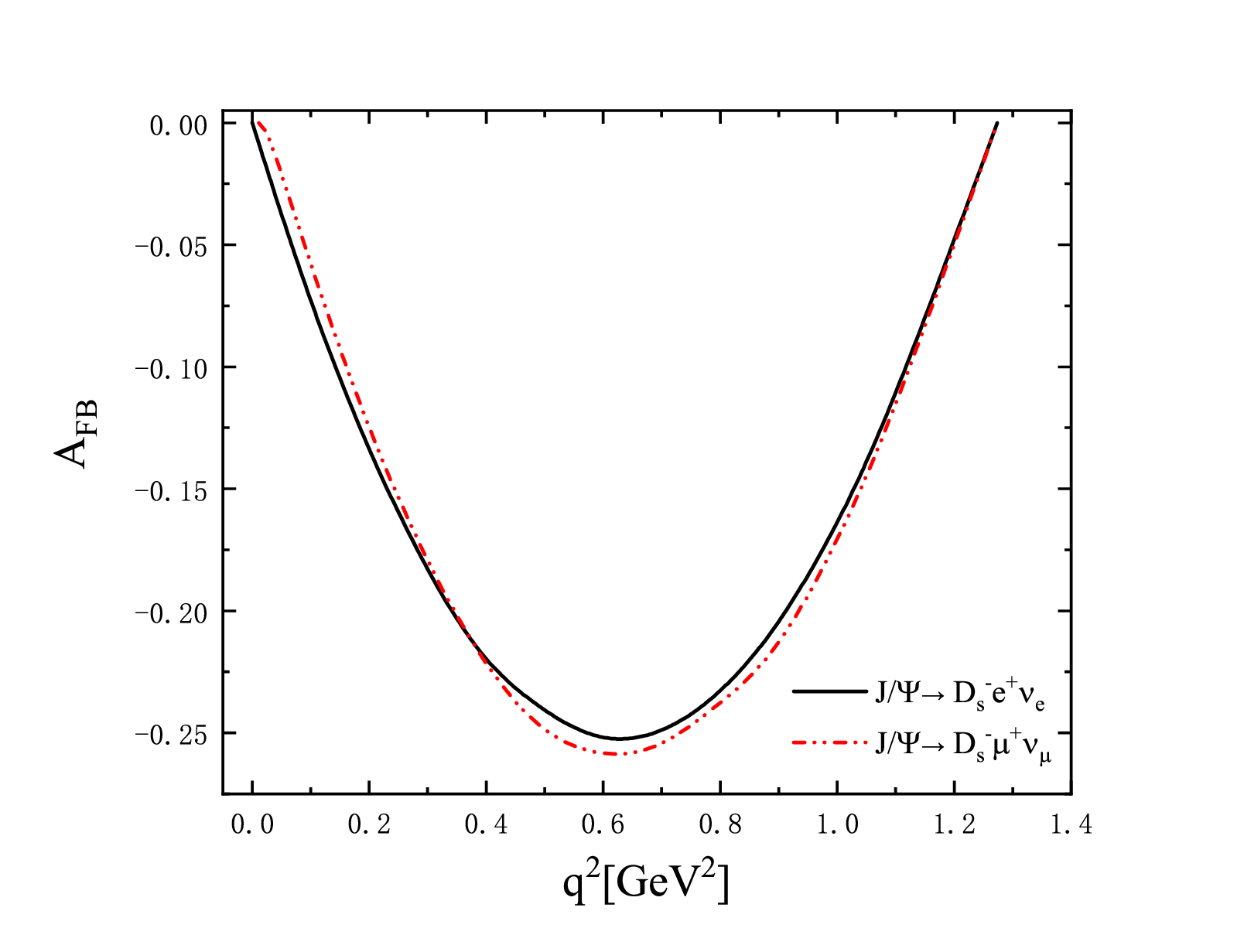}\quad}
  \subfigure[]{\includegraphics[width=0.22\textwidth]{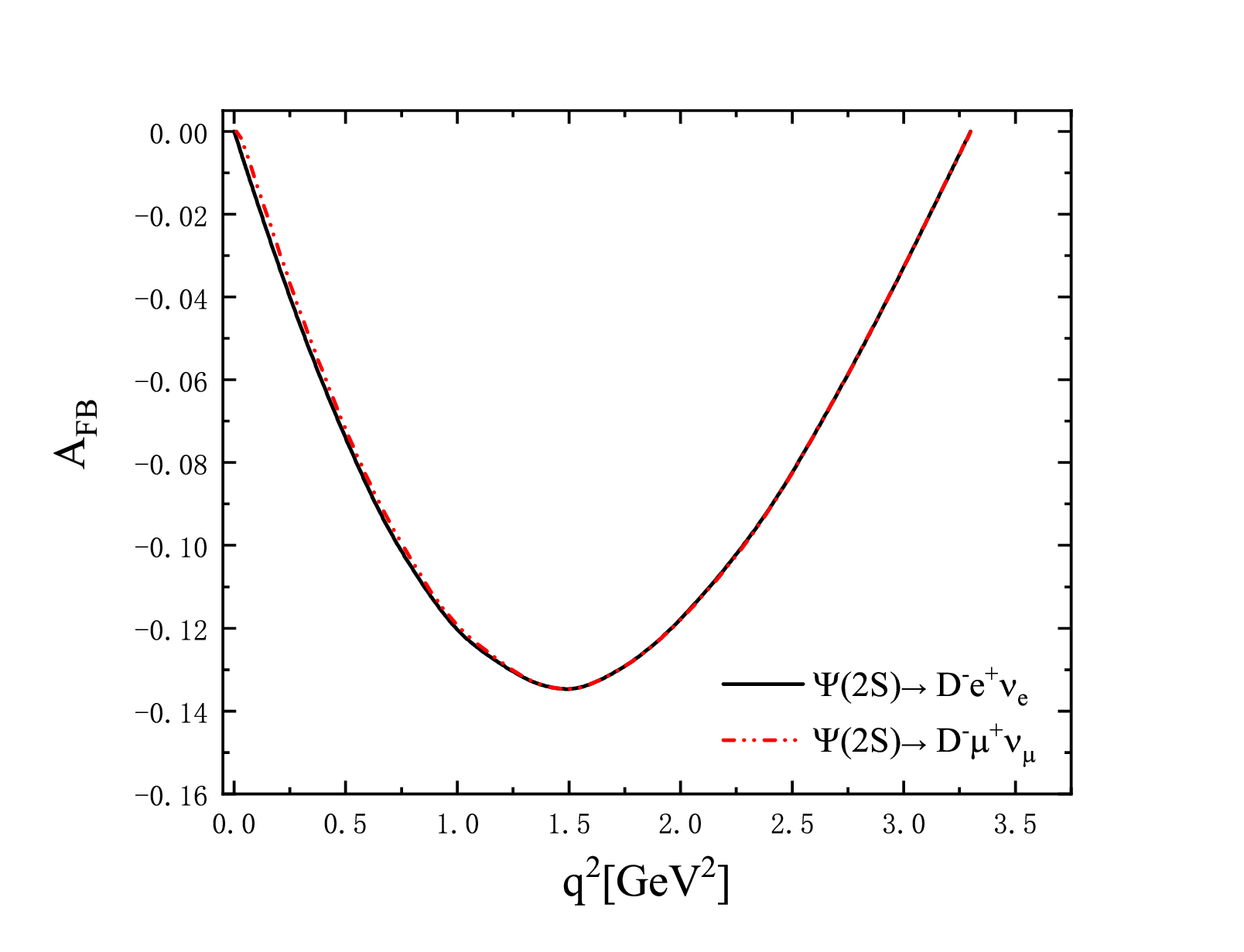}\quad}
  \subfigure[]{\includegraphics[width=0.22\textwidth]{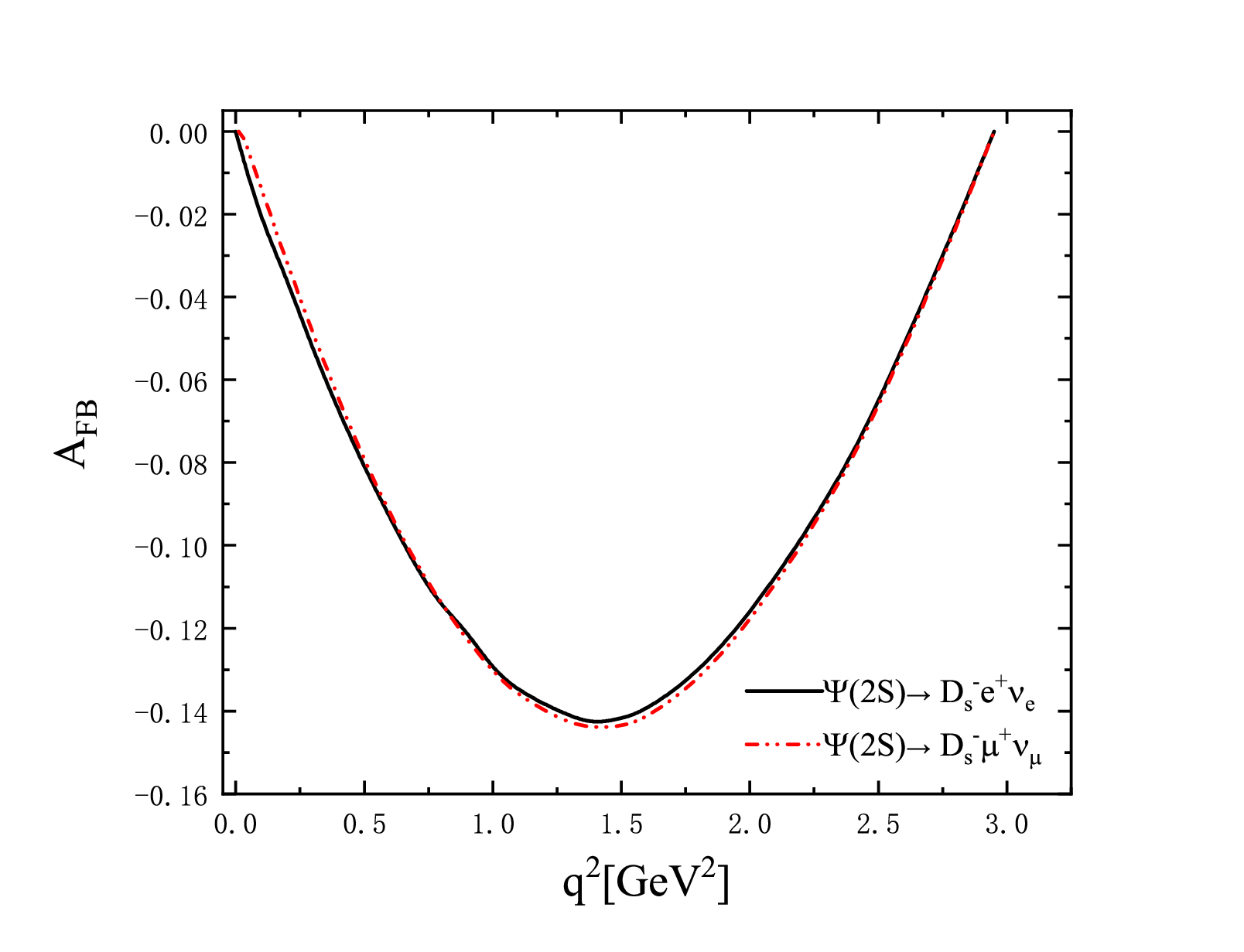}}

\caption{The theoretical predictions for the $q^{2}$  dependences of the forward-backward asymmetries $A_{FB}$.}\label{fig:T4}
\end{figure}
\subsection{Nonleptonic decays}
The decays rates of the charmonium weak decays $\eta_{c}(1S,2S)\to D_{(s)}M$ and $\psi(1S,2S)\to D_{(s)}M$ with
$M$ standing for a  pseudoscalar meson $(P)$ or a vector meson $(V)$ can be written as
\be
\mathcal{B} r\left(\eta_{c}(1S,2S)\to D_{(s)}M\right)=\frac{p_{\mathrm{cm}}}{4 \pi m_{\eta_{c}(1S,2S)}^{2} \Gamma_{\eta_{c}(1S,2S)}}\left|\mathcal{A}\left(\eta_{c}(1S,2S) \to D_{(s)} M\right)\right|^{2}, \\
\mathcal{B} r\left( \psi(1S,2S) \to D_{(s)} M\right)=\frac{p_{\mathrm{cm}}}{12 \pi m_{ \psi(1S,2S)}^{2} \Gamma_{ \psi(1S,2S)}}|A(\psi(1S,2S) \to D_{(s)} M)|^{2}.
\en
where $p_{cm}$ represents the three-momentum of the final  meson $D_{(s)}$ in the rest frame of $\eta_{c}(1S,2S)$ and $\psi(1S,2S)$.

In Tables \ref{eJ} and \ref{eP}, we list the branching ratios of the nonleptonic decays $\eta_{c}(1S,2S)\to D_{(s)}M$ and $ \psi(1S,2S)\to D_{(s)}M$, including the values obtained from Refs. \cite{k.R,R.R,Sun:2015nra,Y.M,Sun:2016ppe,Yang:2016gnh,Sun:2015bxp,Shen:2008zzb,Wang:2007ys,Sun:2015nra} and BESIII  collaboration \cite{BES:2007hqc,BESIII:2014xbo} for comparison, where the uncertainties of our results arise from the full widths of the charmonia $\eta_{c}(1S,2S), \psi(1S,2S)$, the decay constants of the initial and final state mesons, respectively. The decay modes considered here are dominated by the color-favored factorizable contributions and insensitive to the nonfactorizable contributions. Therefore, even with different phenomenological models, the branching ratios for a given decay process of $\eta_{c}(1S,2S)\to D_{(s)}M$ and $ \psi(1S,2S)\to D_{(s)}M$ have the same order of magnitude in many cases. Numerically, we adopt the wilson coefficient $a_{1}=1.26$. The following are some comments:
\begin{enumerate}
\item
The branching ratios for the weak decays $\eta_{c}(2S)\to D_{(s)}M$  are approximately $1.5\sim3$ times larger than those for the corresponding decays $\eta_{c}\to D_{(s)}M$ due to the smaller decay width of $\eta_{c}(2S)$, $\Gamma_{\eta_c(2S)}=(11.3^{+3.2}_{-2.9})$ MeV compared to $\Gamma_{\eta_c}=(32.0\pm0.7)$MeV for $\eta_{c}$ meson, and the larger phase space for $\eta_{c}(2S)$. It is contrary for the cases of weak decays between $\psi(2S)\to D_{(s)}M$ and $J/\Psi\to D_{(s)}M$, where the branching ratios of the latter are about $2\sim3$ times larger than those of the former, because the decay width of $J/\psi$, $\Gamma_{J/\psi}=(92.6\pm1.7)$ keV is only about one-third of that for $\psi(2S)$, $\Gamma_{\psi(2S)}=(294\pm8)$ keV. These numerical relations are similar with those given by the NRQCD \cite{Sun:2015bxp} for the $J/\Psi$ and $\psi(2S)$ decays, while are different for the $\eta_{c}$ and $\eta_{c}(2S)$ decays, where the differences are five times even more large.
\item It is worth mentioning that the branching ratios of the decays $\eta_{c}\to D_{(s)}M$ are in agreement with the results obtained the NRQCD approach\cite{Sun:2015bxp}, while there exists about $2\sim3$ times even more large difference for those of the decays $\eta_{c}(2S)\to D_{(s)}M$.
For the $J/\Psi$ weak decays, the branching ratios of the channels $J/\psi\to D_{(s)}V$ are consistent with most of other theoretical results, such as the NRQCD \cite{Sun:2015bxp}, the BSW model with the parameter $\omega=0.5$ GeV \cite{k.R}, the QCDF approach \cite{Sun:2015nra} and the PQCD approach \cite{Yang:2016gnh}, but are larger than those given in the QCDSR \cite{Y.M} except that of the decay $J/\psi\to D_{(s)}K^{*+}$. While the branching ratios of the decays
 $J/\psi\to D_{(s)}P$ are in agreement with than the calculations given in the BSW model with the parameter $\omega=0.4$ GeV \cite{R.R}, the PQCD approach \cite{Sun:2016ppe} and the
 QCDF approach \cite{Sun:2015nra}, while are smaller than those given in the NRQCD \cite{Sun:2015bxp}.
  As to the $\psi(2S)$  decys, it is similar with the cases of the $J/\Psi$ decays, that is the branching ratios of the decays $\psi(2S)\to D_{(s)}V$
  are comparable with the current only available theoretical results in NRQCD \cite{Sun:2015bxp}, but those of the decays $\psi(2S)\to D_{(s)}P$ are much smaller. Certainly, the decays $J/\psi\to D^{-}_{s}\pi^{+}$, $J/\psi\to D^{-}\pi^{+}$, $J/\psi\to D^{-}_{s}\rho^{+}$ have been detected by the BESIII Collaboration but only with upper bounds \cite{BES:2007hqc,BESIII:2014xbo} being available, which are much above all the theoretical predictions.
\begin{table}[H]
\caption{Branching ratios of the nonleptonic ground charmonium state ($J/\psi, \eta_c$) decays.}
\begin{center}
\scalebox{0.7}{
\begin{tabular}{|c|c|c|c|c|}
\hline\hline
$$&$10^{-12}\times \mathcal{B} r(\eta_{c}\to D^{-}_{s} \pi^{+})$&$10^{-13} \times \mathcal{B} r(\eta_{c}\to D^{-}_{s} K^{+})$&$10^{-13} \times \mathcal{B} r(\eta_{c}\to D^{-} \pi^{+})$&$10^{-14} \times \mathcal{B} r(\eta_{c}\to D^{-} K^{+})$\\
\hline
This work&$6.65^{+0.15+0.01+0.32}_{-0.14-0.01-0.10}$&$4.22^{+0.09+0.00+0.35}_{-0.09-0.00-0.79}$&$3.34^{+0.07+0.01+0.28}_{-0.07-0.02-0.33}$&$2.33^{+0.05+0.01+0.21}_{-0.05-0.01-0.24}$\\
\hline
\cite{Sun:2015bxp}&$7.35$&$4.97$&$4.39$&$3.04$\\
\hline
$$&$10^{-12} \times \mathcal{B} r(\eta_{c}\to D^{-}_{s} \rho^{+})$&$10^{-13} \times \mathcal{B} r(\eta_{c}\to D^{-}_{s} K^{*+})$&$10^{-13} \times \mathcal{B} r(\eta_{c}\to D^{-} \rho^{+})$&$10^{-14} \times \mathcal{B} r(\eta_{c}\to D^{-} K^{*+})$\\
\hline
This work&$6.62^{+0.15+0.03+0.69}_{-0.14-0.03-1.34}$&$3.31^{+0.07+0.02+0.40}_{-0.07-0.02-0.72}$&$3.01^{+0.07+0.00+0.33}_{-0.06-0.00-0.37}$&$1.68^{+0.04+0.00+0.20}_{-0.04-0.00-0.22}$\\
\hline
\cite{Sun:2015bxp}&$5.28$&$1.18$&$4.32$&$1.38$\\
  \hline\hline
  $$&$10^{-10}\times \mathcal{B} r(J/\psi\to D^{-}_{s} \pi^{+})$&$10^{-11} \times \mathcal{B} r(J/\psi\to D^{-}_{s} K^{+})$&$10^{-11} \times \mathcal{B} r(J/\psi\to D^{-} \pi^{+})$&$10^{-12} \times \mathcal{B} r(J/\psi\to D^{-} K^{+})$\\
  \hline
  This work&$3.64^{+0.06+0.34+0.78}_{-0.06-0.38-0.96}$&$2.02^{+0.04+0.18+0.36}_{-0.04-0.20-0.48}$&$1.90^{+0.04+0.17+0.11}_{-0.03-0.19-0.14}$&$1.16^{+0.02+0.03+0.13}_{-0.02-0.02-0.18}$\\
\hline
  \cite{Sun:2015bxp}&$10.9$&$6.18$&$6.37$&$3.79$\\
   \hline
  \cite{Sun:2016ppe}&$4.30$&$2.69$&$2.09$&$1.34$\\
  \hline
  \cite{R.R} \footnotemark[1]&$3.32$&$2.4$&$1.5$&$1.2$\\
  \hline
  \cite{Shen:2008zzb}&$2.5$&$-$&$-$&$50$\\
   \hline
  \cite{Y.M}&$2.0$&$1.6$&$0.80$&$36$\\
  \hline
  \cite{k.R} \footnotemark[2] &$8.74$&$5.5$&$5.5$&$-$\\
  \hline
  \cite{Sun:2015nra}&$4.10$&$2.32$&$2.21$&$1.31$\\
   \hline
  BES\cite{BES:2007hqc}&$<1.4\times10^{6}$&$-$&$<7.5\times10^{6}$&$-$\\
  \hline\hline
  $$&$10^{-9} \times \mathcal{B} r(J/\psi\to D^{-}_{s} \rho^{+})$&$10^{-10} \times \mathcal{B} r(J/\psi\to D^{-}_{s} K^{*+})$&$10^{-10} \times \mathcal{B} r(J/\psi\to D^{-} \rho^{+})$&$10^{-12} \times \mathcal{B} r(J/\psi\to D^{-} K^{*+})$\\
  \hline
  This work&$2.95^{+0.06+0.11+0.15}_{-0.05-0.14-0.19}$&$1.42^{+0.03+0.06+0.07}_{-0.03-0.07-0.10}$&$1.70^{+0.03+0.03+0.07}_{-0.03-0.05-0.10}$&$8.59^{+0.16+0.20+0.42}_{-0.15-0.29-0.60}$\\
  \hline
  \cite{Sun:2015bxp}&$3.82$&$2.00$&$2.12$&$11.4$\\
   \hline
  \cite{R.R}\footnotemark[1]&$1.77$&$0.97$&$0.72$&$4.2$\\
  \hline
  \cite{Shen:2008zzb}&$2.8$&$-$&$-$&$550$\\
  \hline
  \cite{Y.M}&$1.26$&$0.82$&$0.42$&$154$\\
  \hline
  \cite{k.R}\footnotemark[2] &$3.63$&$2.12$&$2.20$&$-$\\
  \hline
  \cite{Sun:2015nra}&$2.21$&$1.22$&$1.09$&$6.14$\\
  \hline
  \cite{Yang:2016gnh}&$3.33$&$1.86$&$1.32$&$8.0$\\
\hline
  BES\cite{BESIII:2014xbo}&$<1.3\times10^{4}$&$-$&$-$&$-$\\
  \hline\hline
\end{tabular}}\label{eJ}
\end{center}
{\footnotesize $^1$ The branching ratios are computed with the average transverse quark momentum $\omega=0.4$ GeV under the WSB model. \\
\footnotesize $^2$ The branching ratios are computed with the average transverse quark momentum $\omega=0.5$ GeV under the WSB model.}
\end{table}

\item Whether the ground or radially excited charmonium state decays, there exists a clear hierarchical pattern among their branching ratios
\begin{footnotesize}
\begin{equation}
\begin{aligned}
\mathcal{B} r\left(\eta_{c}(nS) \rightarrow D^{-}_{s} \pi^{+}\right) &\gg \mathcal{B} r\left(\eta_{c}(nS) \rightarrow D^{-}_{s} K^{+}\right) \sim \mathcal{B} r\left(\eta_{c}(nS) \rightarrow D^{-} \pi^{+}\right) \gg \mathcal{B} r\left(\eta_{c}(nS) \rightarrow D^{-} K^{+}\right),\\
\mathcal{B} r\left(\psi(nS) \rightarrow D^{-}_{s} \pi^{+}\right) &\gg \mathcal{B} r\left(\psi(nS) \rightarrow D^{-}_{s} K^{+}\right) \sim \mathcal{B} r\left(\psi(nS) \rightarrow D^{-} \pi^{+}\right) \gg \mathcal{B} r\left(\psi(nS) \rightarrow D^{-} K^{+}\right),\\
\mathcal{B} r\left(\eta_{c}(nS) \rightarrow D^{-}_{s} \rho^{+}\right) &\gg \mathcal{B} r\left(\eta_{c}(nS) \rightarrow D^{-}_{s} K^{*+}\right) \sim \mathcal{B} r\left(\eta_{c}(nS) \rightarrow D^{-} \rho^{+}\right) \gg \mathcal{B} r\left(\eta_{c}(nS) \rightarrow D^{-} K^{*+}\right),\\
\mathcal{B} r\left(\psi(nS) \rightarrow D^{-}_{s} \rho^{+}\right) &\gg \mathcal{B} r\left(\psi(nS) \rightarrow D^{-}_{s} K^{*+}\right) \sim \mathcal{B} r\left(\psi(nS) \rightarrow D^{-} \rho^{+}\right) \gg \mathcal{B} r\left(\psi(nS) \rightarrow D^{-} K^{*+}\right),
\end{aligned}
\end{equation}
\end{footnotesize}
 which are primarily due to the hierarchical structures of CKM factors $V_{cs}V_{ud}(0.949)\gg V_{cs}V_{us}(0.219)\sim  V_{cd}V_{ud}(0.215)\gg V_{cd}V_{us}(0.049)$.

\begin{table}[H]
\caption{Branching ratios of the nonleptonic radially excited charmonium ($\eta_c(2S), \psi(2S)$) decays.}
\begin{center}
\scalebox{0.7}{
\begin{tabular}{|c|c|c|c|c|}
\hline\hline
$$&$10^{-11}\times \mathcal{B} r(\eta_{c}(2S)\to D^{-}_{s} \pi^{+})$&$10^{-12} \times \mathcal{B} r(\eta_{c}(2S)\to D^{-}_{s} K^{+})$&$10^{-13} \times \mathcal{B} r(\eta_{c}(2S)\to D^{-} \pi^{+})$&$10^{-14} \times \mathcal{B} r(\eta_{c}(2S)\to D^{-} K^{+})$\\
\hline
This work&$1.92^{+0.66+0.24+0.18}_{-0.42-0.44-0.54}$&$1.29^{+0.45+0.13+0.17}_{-0.29-0.33-0.40}$&$7.67^{+2.65+0.78+1.24}_{-1.69-3.29-1.40}$&$5.08^{+1.75+0.63+0.90}_{-1.12-2.24-0.97}$\\
\hline
\cite{Sun:2015bxp}&$3.90$&$2.87$&$21.3$&$15.8$\\
\hline\hline
$$&$10^{-11} \times \mathcal{B} r(\eta_{c}(2S)\to D^{-}_{s} \rho^{+})$&$10^{-13} \times \mathcal{B} r(\eta_{c}(2S)\to D^{-}_{s} K^{*+})$&$10^{-13} \times \mathcal{B} r(\eta_{c}(2S)\to D^{-} \rho^{+})$&$10^{-14} \times \mathcal{B} r(\eta_{c}(2S)\to D^{-} K^{*+})$\\
\hline
This work&$1.92^{+0.66+0.10+0.67}_{-0.42-0.11-1.16}$&$7.05^{+2.43+1.12+1.21}_{-1.56-1.15-2.13}$&$3.93^{+1.36+0.08+0.68}_{-0.87-0.33-0.71}$&$2.54^{+0.88+0.38+0.44}_{-0.56-0.78-0.45}$\\
\hline
\cite{Sun:2015bxp}&$7.24$&$34.7$&$41.3$&$20.2$\\
  \hline\hline
  $$&$10^{-10}\times \mathcal{B} r(\psi(2S)\to D^{-}_{s} \pi^{+})$&$10^{-12} \times \mathcal{B} r(\psi(2S)\to D^{-}_{s} K^{+})$&$10^{-12} \times \mathcal{B} r(\psi(2S)\to D^{-} \pi^{+})$&$10^{-13} \times \mathcal{B} r(\psi(2S)\to D^{-} K^{+})$\\
  \hline
     This work&$1.23^{+0.03+0.08+0.59}_{-0.03-0.18-0.51}$&$8.20^{+0.22+2.62+3.34}_{-0.22-3.50-3.20}$&$7.58^{+0.22+2.32+1.06}_{-0.20-3.40-1.12}$&$4.96^{+0.14+1.42+0.56}_{-0.14-2.14-0.64}$\\
  \hline
   \cite{Sun:2015bxp}&$5.07$&$34.3$&$27.6$&$19$\\
  \hline\hline
  $$&$10^{-9} \times \mathcal{B} r(\psi(2S)\to D^{-}_{s} \rho^{+})$&$10^{-11} \times \mathcal{B} r(\psi(2S)\to D^{-}_{s} K^{*+})$&$10^{-11} \times \mathcal{B} r(\psi(2S)\to D^{-} \rho^{+})$&$10^{-12} \times \mathcal{B} r(\psi(2S)\to D^{-} K^{*+})$\\
  \hline
  This work&$1.22^{+0.03+0.01+0.41}_{-0.03-0.10-0.19}$&$7.31^{+0.20+0.06+0.42}_{-0.19-0.21-0.13}$&$5.55^{+0.16+0.59+0.52}_{-0.15-0.82-0.58}$&$3.31^{+0.09+0.33+0.34}_{-0.09-0.49-0.37}$\\
  \hline
  \cite{Sun:2015bxp}&$1.67$&$9.6$&$8.99$&$5.2$\\
  \hline\hline
\end{tabular}}\label{eP}
\end{center}
\end{table}
\item Our primary focus lies on the nonleptonic decay channels that are most likely to be observed in future collider experiments. In Table \ref{CFS}, we list the ranges of the branching ratios for the Cabibbo-favored decays $\eta_{c}(1S,2S)\to D^{-}_{s} \pi^{+}$, $\psi(1S,2S)\to D^{-}_{s} \pi^{+}$, $\eta_{c}(1S,2S) \to D^{-}_{s} \rho^{+}$, $\psi(1S,2S) \to D^{-}_{s} \rho^{+}$  and the Cabibbo-suppressed decays $\eta_{c} (1S,2S) \to D^{-} K^{+}$, $ \psi(1S,2S) \to D^{-} K^{+}$, $\eta_{c} (1S,2S) \to D^{-} K^{*+}$, $\psi(1S,2S) \to D^{-} K^{*+}$. It is obvious that the decays $\psi(1S,2S) \rightarrow D^{-}_{s} \rho^{+}$ have the largest branching ratios and are most likely to be observed.
    \begin{table}[H]
\caption{The units of the branching ratios of the Cabbibo-favored and Cabibbo-suppressed decay channels.}
\begin{center}
\scalebox{1}{
\begin{tabular}{|c|c|c|c|}
\hline\hline
 Cabbibo-favored decay modes&Units&Cabibbo-suppressed decay modes&Units\\
\hline\hline
$\eta_{c}(1S,2S) \rightarrow D^{-}_{s} \pi^{+}$&$(10^{-12}-10^{-11})$&$\eta_{c}(1S,2S) \rightarrow D^{-} K^{+}$&$10^{-14}$\\
$\psi(1S,2S) \rightarrow D^{-}_{s} \pi^{+}$&$(10^{-11}-10^{-10})$&$\psi(1S,2S) \rightarrow D^{-} K^{+}$&$10^{-13}$\\
$\eta_{c}(1S,2S) \rightarrow D^{-}_{s} \rho^{+} $&$(10^{-12}-10^{-11})$&$\eta_{c} (1S,2S) \rightarrow D^{-} K^{\ast+}$&$10^{-14}$\\
$\psi(1S,2S) \rightarrow D^{-}_{s} \rho^{+}$&$10^{-9}$&$\psi(1S,2S) \rightarrow D^{-} K^{*+}$&$10^{-12}$\\
\hline\hline
\end{tabular}\label{CFS}}
\end{center}
\end{table}
    \item From our calculations, one can obtain the following relative ratios of the branching fractions where the uncertainties from the transition form factors are cancelled
 \begin{footnotesize}
\begin{equation}
\begin{aligned}
R_{\eta_{c}}^{D_s}&\equiv\frac{\mathcal{B} r(\eta_{c}\to D^{-}_{s} K^{+})}{\mathcal{B} r(\eta_{c}\to D^{-}_{s} \pi^{+})}=0.063\pm0.012,\;\;\;\;
R_{\eta_{c}(2S)}^{D_s}&\equiv\frac{\mathcal{B} r(\eta_{c}(2S)\to D^{-}_{s} K^{+})}{\mathcal{B} r(\eta_{c}(2S)\to D^{-}_{s} \pi^{+})}=0.067\pm0.033,\\
R_{J/\psi}^{D_s}&\equiv\frac{\mathcal{B} r(J/\psi\to D^{-}_{s} K^{+})}{\mathcal{B} r(J/\psi\to D^{-}_{s} \pi^{+})}=0.055\pm0.020,\;\;\;\;
R_{\psi(2S)}^{D_s}&\equiv\frac{\mathcal{B} r(\psi(2S)\to D^{-}_{s} K^{+})}{\mathcal{B} r(\psi(2S)\to D^{-}_{s} \pi^{+})}=0.066\pm0.043.\\
\end{aligned}
\end{equation}
\end{footnotesize}
\begin{footnotesize}
\begin{equation}
\begin{aligned}
R_{\eta_{c}}^{D}&\equiv\frac{\mathcal{B} r(\eta_{c}\to D^{-} K^{+})}{\mathcal{B} r(\eta_{c}\to D^{-} \pi^{+})}=0.070\pm0.010,\;\;\;\;
R_{\eta_{c}(2S)}^{D}&\equiv\frac{\mathcal{B} r(\eta_{c}(2S)\to D^{-} K^{+})}{\mathcal{B} r(\eta_{c}(2S)\to D^{-} \pi^{+})}=0.066\pm0.041,\\
R_{J/\psi}^{D}&\equiv\frac{\mathcal{B} r(J/\psi\to D^{-} K^{+})}{\mathcal{B} r(J/\psi\to D^{-} \pi^{+})}=0.061\pm0.011,\;\;\;\;
R_{\psi(2S)}^{D}&\equiv\frac{\mathcal{B} r(\psi(2S)\to D^{-} K^{+})}{\mathcal{B} r(\psi(2S)\to D^{-} \pi^{+})}=0.065\pm0.041,\\
\end{aligned}
\end{equation}
\end{footnotesize}
which are consistent with the estimation $R=\left|V_{u s}\right|^{2} \frac{f_{K}^{2}}{f_{\pi}^{2}}\approx0.074$ obtained from the factorization assumption.  Furthermore, the ratios $R_{J/\psi}^{D_s}$ and $R_{J/\psi}^{D}$ agree well with the results 0.057 and 0.060 given in the QCDF \cite{Sun:2015nra}. Similarly, we can also define the ratios $R_{\eta_c{(1S,2S)}}^\pi,R_{\psi{(1S,2S)}}^\pi$ as follows
\begin{footnotesize}
\begin{equation}
\begin{aligned}
R^{\pi}_{\eta_{c}}&\equiv\frac{\mathcal{B} r(\eta_{c}\to D^{-} \pi^{+})}{\mathcal{B} r(\eta_{c}\to D^{-}_{s} \pi^{+})}=0.050\pm0.006\approx \left|\frac{V_{cd}F^{\eta_{c} D}_{0}(m^{2}_{\pi})}{V_{cs}F^{\eta_{c} D_{s}}_{0}(m^{2}_{\pi})}\right|^{2} =0.041,\\
R^{\pi}_{\eta_{c}(2S)}&\equiv\frac{\mathcal{B} r(\eta_{c}(2S)\to D^{-} \pi^{+})}{\mathcal{B} r(\eta_{c}(2S)\to D^{-}_{s} \pi^{+})}=0.040\pm0.019 \approx\left|\frac{V_{cd}F^{\eta_{c} (2S) D}_{0}(m^{2}_{\pi})}{V_{cs}F^{\eta_{c}(2S) D_{s}}_{0}(m^{2}_{\pi})}\right|^{2}=0.035,\\
R^{\pi}_{J/\psi}&\equiv\frac{\mathcal{B} r(J/\psi\to D^{-} \pi^{+})}{\mathcal{B} r(J/\psi\to D^{-}_{s} \pi^{+})}=0.052\pm0.015\approx\left|\frac{V_{cd}A^{J/\psi D}_{0}(m^{2}_{\pi})}{V_{cs}A^{J/\psi D_{s}}_{0}(m^{2}_{\pi})}\right|^{2}=0.043,\\
R^{\pi}_{\psi(2S)}&\equiv\frac{\mathcal{B} r(\psi(2S)\to D^{-} \pi^{+})}{\mathcal{B} r(\psi(2S)\to D^{-}_{s} \pi^{+})}=0.061\pm0.040\approx\left|\frac{V_{cd}A^{\psi(2S) D}_{0}(m^{2}_{\pi})}{V_{cs}A^{\psi(2S) D_{s}}_{0}(m^{2}_{\pi})}\right|^{2}=0.055.\\
\end{aligned}
\end{equation}
\end{footnotesize}
\end{enumerate}
\section{Summary}\label{sum}
The charmonium weak decays provide a unique perspective on the underlying structures and dynamical mechanisms of hadrons and currents. With the anticipation of abundant data samples on charmonium at high-luminosity heavy-flavor experiments, we calculated some semileptonic and nonleptonic weak decays of charmonia $\eta_{c}(1S,2S)$ and $\psi(1S,2S)$ using the covariant light-front quark model. Here the nonperturbative weak transition form factors play a crucial role in evaluating the weak meson decay amplitudes. We extended analytically the expressions of the form factors of the transitions $\eta_{c}(1S,2S)\to D_{(s)}$ and $\psi(1S,2S)\to D_{(s)}$  in the space-like region to the time-like region using the double-pole model.  The following are some points
\begin{enumerate}
\item In our considered decays, the channels $J/\psi\to D_{s}^{-}\ell^{+}\nu_{\ell}$ and $J/\psi\to D^{-}_{s}\rho^{+}$ have the largest branching ratios, which are very close to or even upto $10^{-9}$. These values are still much below the present experimental upper bounds.
\item The branching ratios for the semileptonic decays $J/\Psi\to D^{-}_{(s)}\ell^+\nu_{\ell}$ are well consistent the results obtained from the BSW model, but some three or more times as large as those given by the BS
approach, the CCQM and the QCDSM. The semileptonic decays of the radially excited charmonia $\psi(2S)$ and $\eta_c(2S)$ have not been studied by any other theory. We find that $Br(J/\Psi\to D^{-}_{(s)}\ell^+\nu_{\ell})$ are about three orders of magnitude larger than $Br(\eta_c\to D^{-}_{(s)}\ell^+\nu_{\ell})$, and $Br(\psi(2S)\to D^{-}_{(s)}\ell^+\nu_{\ell})$ are about two orders of magnitude larger than $Br(\eta_{c}(2S)\to D^{-}_{(s)}\ell^+\nu_{\ell})$.

\item Whether the semileptonic or the nonleptonic charmonium weak decays, the branching ratios for the ground state $\eta_c$ decays are smaller than those of the radially excited state $\eta_c(2S)$ ones. It is contrary to the cases of $\psi(1S,2S)$ decays, where the branching ratios of the ground sate $J/\Psi$ decays are larger than those of the radially excited state $\psi(2S)$ ones. It is because of the larger (smaller) decay width of $\eta_c$ $(J/\Psi)$ compared with that of its radially excited state.
\item The ratios of the forward-backward asymmetries $A_{FB}^\mu/A_{FB}^e$ between the semileptonic decays
 $\eta_c(1S,2S)\to D^-_{(s)}\mu^+\nu_{\mu}$ and $\eta_c(1S,2S)\to D^-_{(s)}e^+\nu_{e}$ are in the order of $10^{4}$, and the forward-backward asymmetries for the decays $\psi(1S,2S)\to D^-_{(s)}\mu^+\nu_{\mu}$ and $\psi(1S,2S)\to D^-_{(s)}e^+\nu_{e}$ become minus in sign and lies in the range of $(-0.3\sim-0.2)$.
\item The longitudinal polarization fractions $f_L$ are close to each other between the decays $\psi(1S,2S) \to D^{-}_{(s)}e^{+}\nu_{e}$ and
$\psi(1S,2S) \to D^{-}_{(s)}\mu^{+}\nu_{\mu}$. Furthermore, the longitudinal and transverse polarization fractions for each decay are comparable.
\item Whether the ground or radially excited charmonium decays, the final states $D^-_s\pi^{+}(D^-_s\rho^+)$ always own the largest yield and $D^-K^+(D^-K^{*+})$ always have the smallest production, which are connected with the
   hierarchical structures of CKM factors, $V_{cs}V_{ud}(0.949)\gg V_{cd}V_{us}(0.049)$.
\end{enumerate}
\section*{Acknowledgment}
This work is partly supported by the National Natural Science
Foundation of China under Grant No. 11347030, by the Program of
Science and Technology Innovation Talents in Universities of Henan
Province 14HASTIT037, and the Natural Science Foundation of Henan
Province under grant no. 232300420116..
\appendix
\section{Some specific rules under the $p^-$ intergration}
When preforming the integraion, we need to include the zero-mode contribution. It amounts to performing the integration in a proper way in the CLFQM. Specificlly we
use the following rules given in Refs. \cite{jaus,Y. Cheng}
\be
\hat{p}_{1 \mu}^{\prime} &\doteq &   P_{\mu}
A_{1}^{(1)}+q_{\mu} A_{2}^{(1)},\\
\hat{p}_{1 \mu}^{\prime}
\hat{p}_{1 \nu}^{\prime}  &\doteq & g_{\mu \nu} A_{1}^{(2)} +P_{\mu}
P_{\nu} A_{2}^{(2)}+\left(P_{\mu} q_{\nu}+q_{\mu} P_{\nu}\right)
A_{3}^{(2)}+q_{\mu} q_{\nu} A_{4}^{(2)},\\
Z_{2}&=&\hat{N}_{1}^{\prime}+m_{1}^{\prime 2}-m_{2}^{2}+\left(1-2
x_{1}\right) M^{\prime 2} +\left(q^{2}+q \cdot P\right)
\frac{p_{\perp}^{\prime} \cdot q_{\perp}}{q^{2}},\\
\hat{p}_{1 \mu}^{\prime} \hat{N}_{2} & \rightarrow & q_{\mu}\left[A_{2}^{(1)} Z_{2}+\frac{q \cdot P}{q^{2}} A_{1}^{(2)}\right],
\en
\be
\hat{p}_{1 \mu}^{\prime} \hat{p}_{1 \nu}^{\prime} \hat{N}_{2} & \rightarrow &g_{\mu \nu} A_{1}^{(2)} Z_{2}+q_{\mu} q_{\nu}\left[A_{4}^{(2)} Z_{2}+2 \frac{q \cdot P}{q^{2}} A_{2}^{(1)} A_{1}^{(2)}\right],\\
A_{1}^{(1)}&=&\frac{x_{1}}{2}, \quad A_{2}^{(1)}=
A_{1}^{(1)}-\frac{p_{\perp}^{\prime} \cdot q_{\perp}}{q^{2}},\quad A_{3}^{(2)}=A_{1}^{(1)} A_{2}^{(1)},\\
A_{4}^{(2)}&=&\left(A_{2}^{(1)}\right)^{2}-\frac{1}{q^{2}}A_{1}^{(2)},\quad A_{1}^{(2)}=-p_{\perp}^{\prime 2}-\frac{\left(p_{\perp}^{\prime}
\cdot q_{\perp}\right)^{2}}{q^{2}}, \quad A_{2}^{(2)}=\left(A_{1}^{(1)}\right)^{2}.  \en
\section{EXPRESSIONS OF $\eta_{c}(\psi) \rightarrow D_{(s)}$ FORM FACTORS}
\be
S_{\mu}^{\eta_c D_{(s)}}&=&\operatorname{Tr}\left[\gamma_{5}\left(\not p_{1}^{\prime \prime}+m_{1}^{\prime \prime}\right) \gamma_{\mu}\left(\not p_{1}^{\prime}
+m_{1}^{\prime}\right) \gamma_{5}\left(-\not p_{2}+m_{2}\right)\right] \non
&=& 2 p_{1 \mu}^{\prime}\left[M^{\prime 2}+M^{\prime \prime 2}-q^{2}-2 N_{2}-\left(m_{1}^{\prime}-m_{2}\right)^{2}-\left(m_{1}^{\prime \prime}
-m_{2}\right)^{2}+\left(m_{1}^{\prime}-m_{1}^{\prime \prime}\right)^{2}\right]\non
&&+q_{\mu}\left[q^{2}-2 M^{\prime 2}+N_{1}^{\prime}-N_{1}^{\prime \prime}+2 N_{2}+2\left(m_{1}^{\prime}-m_{2}\right)^{2}-\left(m_{1}^{\prime}
-m_{1}^{\prime \prime}\right)^{2}\right] \non
&&+P_{\mu}\left[q^{2}-N_{1}^{\prime}-N_{1}^{\prime \prime}-\left(m_{1}^{\prime}-m_{1}^{\prime \prime}\right)^{2}\right],
\en
\be
S_{\mu \nu}^{\psi D_{(s)}}&=&\left(S_{V}^{\psi D_{(s)} }-S_{A}^{\psi D_{(s)}}\right)_{\mu \nu}\non
&=&\operatorname{Tr}\left[\left(\gamma_{\nu}-\frac{1}{W_{V}^{\prime \prime}}\left(p_{1}^{\prime \prime}-p_{2}\right)_{\nu}\right)\left(p_{1}^{\prime \prime}
+m_{1}^{\prime \prime}\right)\left(\gamma_{\mu}-\gamma_{\mu} \gamma_{5}\right)\left(\not p_{1}^{\prime}+m_{1}^{\prime}\right) \gamma_{5}\left(-\not p_{2}
+m_{2}\right)\right] \non
&=&-2 i \epsilon_{\mu \nu \alpha \beta}\left\{p_{1}^{\prime \alpha} P^{\beta}\left(m_{1}^{\prime \prime}-m_{1}^{\prime}\right)
+p_{1}^{\prime \alpha} q^{\beta}\left(m_{1}^{\prime \prime}+m_{1}^{\prime}-2 m_{2}\right)+q^{\alpha} P^{\beta} m_{1}^{\prime}\right\} \non
&&+\frac{1}{W_{V}^{\prime \prime}}\left(4 p_{1 \nu}^{\prime}-3 q_{\nu}-P_{\nu}\right) i \epsilon_{\mu \alpha \beta \rho} p_{1}^{\prime \alpha} q^{\beta} P^{\rho}\non &&
+2 g_{\mu \nu}\left\{m_{2}\left(q^{2}-N_{1}^{\prime}-N_{1}^{\prime \prime}-m_{1}^{\prime 2}-m_{1}^{\prime \prime 2}\right)
-m_{1}^{\prime}\left(M^{\prime \prime 2}-N_{1}^{\prime \prime}-N_{2}-m_{1}^{\prime \prime 2}-m_{2}^{2}\right)\right.\non
&&\left.-m_{1}^{\prime \prime}\left(M^{\prime 2}-N_{1}^{\prime}-N_{2}-m_{1}^{\prime 2}-m_{2}^{2}\right)
-2 m_{1}^{\prime} m_{1}^{\prime \prime} m_{2}\right\} \nonumber \en
\be
\;\;\;\;\;\;&&+8 p_{1 \mu}^{\prime} p_{1 \nu}^{\prime}\left(m_{2}-m_{1}^{\prime}\right)-2\left(P_{\mu} q_{\nu}
+q_{\mu} P_{\nu}+2 q_{\mu} q_{\nu}\right) m_{1}^{\prime}+2 p_{1 \mu}^{\prime} P_{\nu}\left(m_{1}^{\prime}-m_{1}^{\prime \prime}\right)\non &&
+2 p_{1 \mu}^{\prime} q_{\nu}\left(3 m_{1}^{\prime}-m_{1}^{\prime \prime}-2 m_{2}\right)
+2 P_{\mu} p_{1 \nu}^{\prime}\left(m_{1}^{\prime}+m_{1}^{\prime \prime}\right)+2 q_{\mu} p_{1 \nu}^{\prime}\left(3 m_{1}^{\prime}+m_{1}^{\prime \prime}-2 m_{2}\right)\non &&
+\frac{1}{2 W_{V}^{\prime \prime}}\left(4 p_{1 \nu}^{\prime}-3 q_{\nu}-P_{\nu}\right)\left\{2 p_{1 \mu}^{\prime}\left[M^{\prime 2}
+M^{\prime \prime 2}-q^{2}-2 N_{2}+2\left(m_{1}^{\prime}-m_{2}\right)\left(m_{1}^{\prime \prime}+m_{2}\right)\right]\right.\non&&
+q_{\mu}\left[q^{2}-2 M^{\prime 2}+N_{1}^{\prime}-N_{1}^{\prime \prime}+2 N_{2}-\left(m_{1}^{\prime}+m_{1}^{\prime \prime}\right)^{2}+2\left(m_{1}^{\prime}-m_{2}\right)^{2}\right]\non&&
\left.+P_{\mu}\left[q^{2}-N_{1}^{\prime}-N_{1}^{\prime \prime}-\left(m_{1}^{\prime}+m_{1}^{\prime \prime}\right)^{2}\right]\right\} .
\label{sptov}\en
The following are the analytical expressions of the form factors of transitions $\eta_c(1S, 2S)\to D_{(s)}$, $\psi(1S,2S)\to (1S, 2S)\to D_{(s)}$  in the covariant light-front quark model
\begin{footnotesize}
\begin{eqnarray}
F^{\eta_c D_{(s)}}_{1}\left(q^{2}\right)&=&\frac{N_{c}}{16 \pi^{3}} \int d x_{2} d^{2} p_{\perp}^{\prime} \frac{h_{\eta_c}^{\prime}
h_{D_{(s)}}^{\prime \prime}}{x_{2} \hat{N}_{1}^{\prime} \hat{N}_{1}^{\prime \prime}}\left[x_{1}\left(M_{0}^{\prime 2}+M_{0}^{\prime \prime 2}\right)+x_{2} q^{2}\right.-x_{2}\left(m_{1}^{\prime}-m_{1}^{\prime \prime}\right)^{2}\non
&&-x_{1}\left(m_{1}^{\prime}-m_{2}\right)^{2}-x_{1}\left(m_{1}^{\prime \prime}-m_{2}\right)^{2}]\\
F^{\eta_c D_{(s)}}_{0}\left(q^{2}\right)&=&F^{\eta_c D_{(s)}}_{1}(q^{2})+\frac{q^2}{(q\cdot P)}\frac{N_{c}}{16 \pi^{3}}  \int d x_{2} d^{2} p_{\perp}^{\prime} \frac{2 h_{\eta_c}^{\prime}
h_{D_{(s)}}^{\prime \prime}}{x_{2} \hat{N}_{1}^{\prime} \hat{N}_{1}^{\prime \prime}}\left\{-x_{1} x_{2} M^{\prime 2}
-p_{\perp}^{\prime 2}-m_{1}^{\prime} m_{2}\right.\notag\\
&&+\left(m_{1}^{\prime \prime}-m_{2}\right)\left(x_{2} m_{1}^{\prime}
+x_{1} m_{2}\right)+2 \frac{q \cdot P}{q^{2}}\left(p_{\perp}^{\prime 2}+2 \frac{\left(p_{\perp}^{\prime} \cdot q_{\perp}\right)^{2}}{q^{2}}\right)
+2 \frac{\left(p_{\perp}^{\prime} \cdot q_{\perp}\right)^{2}}{q^{2}}\notag\\
&&\left.-\frac{p_{\perp}^{\prime} \cdot q_{\perp}}{q^{2}}\left[M^{\prime \prime 2}-x_{2}\left(q^{2}+q \cdot P\right)-\left(x_{2}-x_{1}\right) M^{\prime 2}+2 x_{1} M_{0}^{\prime 2}\right.\right.\notag\\
&&\left.\left.-2\left(m_{1}^{\prime}-m_{2}\right)\left(m_{1}^{\prime}+m_{1}^{\prime \prime}\right)\right]\right\},\\
V^{\psi D_{(s)}}(q^{2})&=&\frac{N_{c}(M^{'}+M^{''})}{16 \pi^{3}} \int d x_{2} d^{2} p_{\perp}^{\prime} \frac{2 h_{\psi}^{\prime}
 h_{D_{(s)}}^{\prime \prime}}{x_{2} \hat{N}_{1}^{\prime} \hat{N}_{1}^{\prime \prime}}\left\{x_{2} m_{1}^{\prime}
 +x_{1} m_{2}+\left(m_{1}^{\prime}-m_{1}^{\prime \prime}\right) \frac{p_{\perp}^{\prime} \cdot q_{\perp}}{q^{2}}\right.\non &&\left.
 +\frac{2}{w_{V}^{\prime \prime}}\left[p_{\perp}^{\prime 2}+\frac{\left(p_{\perp}^{\prime} \cdot q_{\perp}\right)^{2}}{q^{2}}\right]\right\},\\
 A_0^{\psi D_{(s)}}(q^{2})&=& \frac{M^{'}+M^{''}}{2M^{''}}A_1^{\psi D_{(s)}}(q^{2})-\frac{M^{'}-M^{''}}{2M^{''}}A_2^{\psi D_{(s)}}(q^{2})-\frac{q^2}{2M^{''}}\frac{N_{c}}{16 \pi^{3}} \int d x_{2} d^{2} p_{\perp}^{\prime} \frac{h_{\psi}^{\prime} h_{D_{(s)}}^{\prime \prime}}{x_{2} \hat{N}_{1}^{\prime}
\hat{N}_{1}^{\prime \prime}}\left\{2\left(2 x_{1}-3\right)\right.\non &&\left.\left(x_{2} m_{1}^{\prime}+x_{1} m_{2}\right)-8\left(m_{1}^{\prime}-m_{2}\right)
\times\left[\frac{p_{\perp}^{\prime 2}}{q^{2}}
+2 \frac{\left(p_{\perp}^{\prime} \cdot q_{\perp}\right)^{2}}{q^{4}}\right]-\left[\left(14-12 x_{1}\right) m_{1}^{\prime}\right.\right. \non &&\left.\left.-2 m_{1}^{\prime \prime}-\left(8-12 x_{1}\right) m_{2}\right] \frac{p_{\perp}^{\prime} \cdot q_{\perp}}{q^{2}}
+\frac{4}{w_{V}^{\prime \prime}}\left(\left[M^{\prime 2}+M^{\prime \prime 2}-q^{2}+2\left(m_{1}^{\prime}-m_{2}\right)\left(m_{1}^{\prime \prime}
+m_{2}\right)\right]\right.\right.\non &&\left.\left.\times\left(A_{3}^{(2)}+A_{4}^{(2)}-A_{2}^{(1)}\right)
+Z_{2}\left(3 A_{2}^{(1)}-2 A_{4}^{(2)}-1\right)+\frac{1}{2}\left[x_{1}\left(q^{2}+q \cdot P\right)
-2 M^{\prime 2}-2 p_{\perp}^{\prime} \cdot q_{\perp}\right.\right.\right.\non &&\left.\left.\left.-2 m_{1}^{\prime}\left(m_{1}^{\prime \prime}+m_{2}\right)
-2 m_{2}\left(m_{1}^{\prime}-m_{2}\right)\right]\left(A_{1}^{(1)}+A_{2}^{(1)}-1\right) q \cdot P\left[\frac{p_{\perp}^{\prime 2}}{q^{2}}
+\frac{\left(p_{\perp}^{\prime} \cdot q_{\perp}\right)^{2}}{q^{4}}\right]\right.\right.\non &&\left.\left.\times\left(4 A_{2}^{(1)}-3\right)\right)\right\},\;\;\;\\
A_1^{\psi D_{(s)}}(q^{2})&=& -\frac{1}{M^{'}+M^{''}}\frac{N_{c}}{16 \pi^{3}} \int d x_{2} d^{2} p_{\perp}^{\prime} \frac{h_{\psi}^{\prime} h_{D_{(s)}}^{\prime \prime}}{x_{2}
\hat{N}_{1}^{\prime}
\hat{N}_{1}^{\prime \prime}}\left\{2 x_{1}\left(m_{2}-m_{1}^{\prime}\right)\left(M_{0}^{\prime 2}+M_{0}^{\prime \prime 2}\right)
-4 x_{1} m_{1}^{\prime \prime} M_{0}^{\prime 2}\right.\non
&&\left.+2 x_{2} m_{1}^{\prime} q \cdot P+2 m_{2} q^{2}-2 x_{1} m_{2}\left(M^{\prime 2}+M^{\prime \prime 2}\right)+2\left(m_{1}^{\prime}-m_{2}\right)\left(m_{1}^{\prime}
+m_{1}^{\prime \prime}\right)^{2}+8\left(m_{1}^{\prime}-m_{2}\right) \right.\non &&
\left. \times\left[p_{\perp}^{\prime 2}+\frac{\left(p_{\perp}^{\prime}
\cdot q_{\perp}\right)^{2}}{q^{2}}\right]+2\left(m_{1}^{\prime}+m_{1}^{\prime \prime}\right)\left(q^{2}+q \cdot P\right) \frac{p_{\perp}^{\prime} \cdot q_{\perp}}{q^{2}}
-4 \frac{q^{2} p_{\perp}^{\prime 2}+\left(p_{\perp}^{\prime} \cdot q_{\perp}\right)^{2}}{q^{2} w_{V}^{\prime \prime}}
\right.\non && \left.\times\left[2 x_{1}\left(M^{\prime 2}+M_{0}^{\prime 2}\right)-q^{2}-q \cdot P-2\left(q^{2}+q \cdot P\right) \frac{p_{\perp}^{\prime} \cdot q_{\perp}}{q^{2}}-2\left(m_{1}^{\prime}-m_{1}^{\prime \prime}\right)\left(m_{1}^{\prime}-m_{2}\right)\right]\right\},\;\;\;\;\;
\en
\be
A_2^{\psi D_{(s)}}(q^{2})&=& \frac{N_{c}(M^{'}+M^{''})}{16 \pi^{3}} \int d x_{2} d^{2} p_{\perp}^{\prime} \frac{2 h_{\psi}^{\prime} h_{D_{(s)}}^{\prime \prime}}{x_{2} \hat{N}_{1}^{\prime}
\hat{N}_{1}^{\prime \prime}}\left\{\left(x_{1}-x_{2}\right)\left(x_{2} m_{1}^{\prime}+x_{1} m_{2}\right)-\frac{p_{\perp}^{\prime} \cdot q_{\perp}}{q^{2}}\left[2 x_{1} m_{2}
+m_{1}^{\prime \prime} \right.\right.\non &&
\left.\left.+\left(x_{2}-x_{1}\right) m_{1}^{\prime}\right]-2 \frac{x_{2} q^{2}+p_{\perp}^{\prime} \cdot q_{\perp}}{x_{2} q^{2} w_{V}^{\prime \prime}}\left[p_{\perp}^{\prime} \cdot p_{\perp}^{\prime \prime}
+\left(x_{1} m_{2}+x_{2} m_{1}^{\prime}\right)\left(x_{1} m_{2}-x_{2} m_{1}^{\prime \prime}\right)\right]\right\}.
\end{eqnarray}
\end{footnotesize}

\end{document}